\documentclass[reprint,superscriptaddress,floatfix,longbibliography,tightenlines,nobibnotes,nofootinbib,amsmath,amssymb,aps,pra]{revtex4-1}

\usepackage{graphicx}
\usepackage{color}
\usepackage{bm}
\usepackage{textcomp}
\usepackage{hyperref}
\hypersetup{colorlinks=true,urlcolor= blue,citecolor=blue,linkcolor= blue,bookmarks=true,bookmarksopen=false}
\usepackage{verbatim}
\usepackage{blindtext}
\usepackage{natbib}
\setcitestyle{numbers,square}

\begin{document}
	
	\title{Transport in armchair graphene nanoribbons and in ordinary waveguides}
	\author{M. Zubair}
	\email{muhammad.zubair@mail.concordia.ca}
	\affiliation{Department of Physics, Concordia University, 7141 Sherbrooke Ouest, Montreal, Quebec H4B 1R6, Canada}
	\author{M. Bahrami}
	\email{mousabahrami@gmail.com}
	\affiliation{Department of Computer Science and Mathematics
		Nippising University
		100 College Drive, North Bay, ON, P1B 8L7, Canada}
	\author{P. Vasilopoulos }
	\email{p.vasilopoulos@concordia.ca}
	\affiliation{Department of Physics, Concordia University, 7141 Sherbrooke Ouest, Montreal, Quebec H4B 1R6, Canada}
	\begin{abstract}
		We study dc and ac transport along armchair graphene nanoribbons %and along ordinary waveguides. In the former case we 
		using the ${\bf k\cdot p}$ spectrum and eigenfunctions  and general linear-response expressions for the conductivities. Then we contrast the results with those for transport along ordinary waveguides. In all cases we
		assess the influence of elastic scattering by impurities, describe it quantitatively with a Drude-type contribution to the current previously not reported, and evaluate the corresponding relaxation time for long- and short-range impurity potentials. We show that this contribution dominates the response at very low frequencies. % $\omega$. 
		In both cases the conductivities increase with the electron density and show cusps when new subbands start being occupied.
		As functions of the frequency the conductivities in armchair graphene nanoribbons exhibit a much richer peak structure than in ordinary waveguides: in the former intraband and interband transitions are allowed whereas in the latter only the intraband ones occur. This difference can be traced to that  between the corresponding spectra and eigenfunctions.
		
	\end{abstract}
	\maketitle
	
	\section{Introduction} 
	
	Graphene nanoribbons have been studied extensively theoretically and experimentally. Previous studies focused on their electronic structure, spectrum, and eigenfunctions \cite{r1}, optical properties \cite{r2, r22}, elementary excitations \cite{r3}, magnetic susceptibility \cite{r4, nr1}, excitonic effects \cite{r5}.  A short review of transport properties, focused on localization concepts, appeared in Ref. \cite{r6}, some numerical results in Ref. \cite{r7},  numerically studied thermal transport  in Ref. \cite{r8}, and spin tranport in substitutionally doped, zig-zag graphene nanoribbons in Ref. \cite{r88}. Experimental results have also been reported \cite{r9}.
	The influence of impurity scattering or disorder though has received a limited attention \cite{r88}. In particular, we are not aware of any study  of dc and ac transport, say, within linear-response theory, that takes into account 
	scattering by randomly distributed impurities, most of the studies use scattering-independent Kubo formulas or consider scattering numerically. 
	
	In this work  we study dc and ac transport along armchair graphene nanoribbons (AGNRs) or
	ordinary waveguides using linear-response, scattering-dependent and scattering-independent expressions for the conductivities. In the former case we evaluate the  relaxation time for long- and short-range impurity potentials. We present the basics in Sec. II and the conductivities in Sec. III. A summary follows in Sec. IV.
	\begin{figure}[t]
		\centering
		%\hspace*{-0.4cm}
		%\vspace*{-2cm}
		\includegraphics[width=8cm, height=3.5cm%width=.9\linewidth
		]{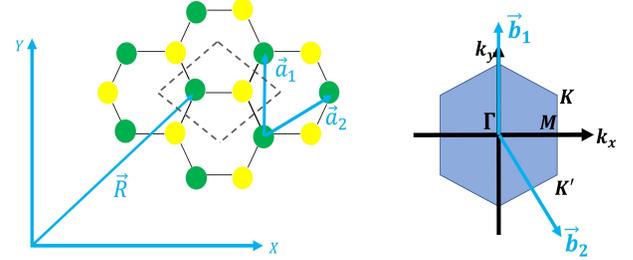}%{Fig1.eps}
		%\includegraphics[width=3cm, height=3cm%width=.9\linewidth
		%]{Fig1b.eps}
		%\vspace{-3cm}
		\caption{Left panel: Graphene unit cell (dashed rhombus) and\\ its  primitive vectors $\vec{a}_1$ and $\vec{a}_2 $.
		 %where the dashed rhombic indicates the unit cell. 
		 Right panel: The corresponding Brillouin zone %of a graphene lattice  
		 with $ \protect  \vec b_{1}  $  and $\protect \vec b_{2}$ the reciprocal lattice vectors. }
		%\caption {Geometry for  an armchair graphene nanoribbon along the $x$ axis. Left panel:    $ \protect  \vec a_{1}  $  and $\protect \vec a_{2}$ are the primitive lattice vectors of graphene. 
		%The rectangular dashed box represents the unit cell of the % graphene
		%%\textcolor{red}{
		%AGNR. %}lattice. %The ribbons are %formed along the $x$ axis. 
		%Right panel: Brillouin zone of a graphene lattice  with $ \protect  \vec b_{1}  $  and $\protect \vec b_{2}$ the reciprocal lattice vectors. }
		\label{f}
	\end{figure}
	
	%%%%%%%
	
	\section{AGNRs, %Nanoribbons, 
		ordinary waveguides}
	\subsection{AGNRs}%rmchair graphene nanoribbons} % of  (AGNRs)}
	
\begin{figure}[t]
		\centering
		\includegraphics[width=8.5cm, height=3.5cm
		]{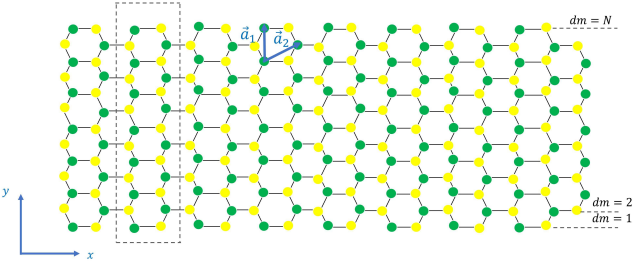}
		\caption{
			Geometry of an AGNR. The dashed box shows the unit cell and $dm$ represents the dimmer number.}
		\label{ribbon}
	\end{figure}

\begin{figure}[t]
		\centering
		\includegraphics[width=4cm, height=4.3cm
		]{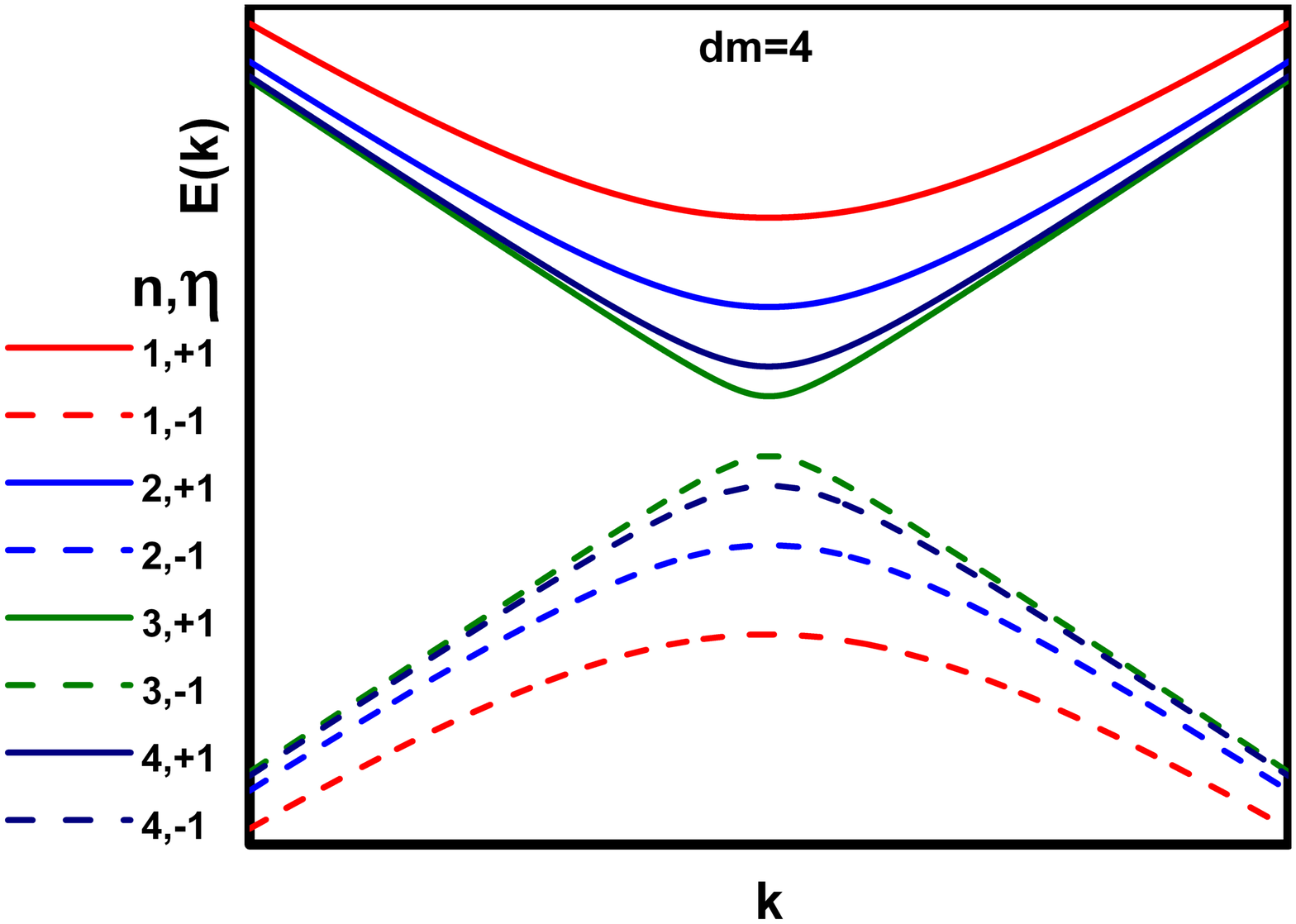}
		\hspace{0.3cm}
		\includegraphics[width=4cm, height=4.3cm
		]{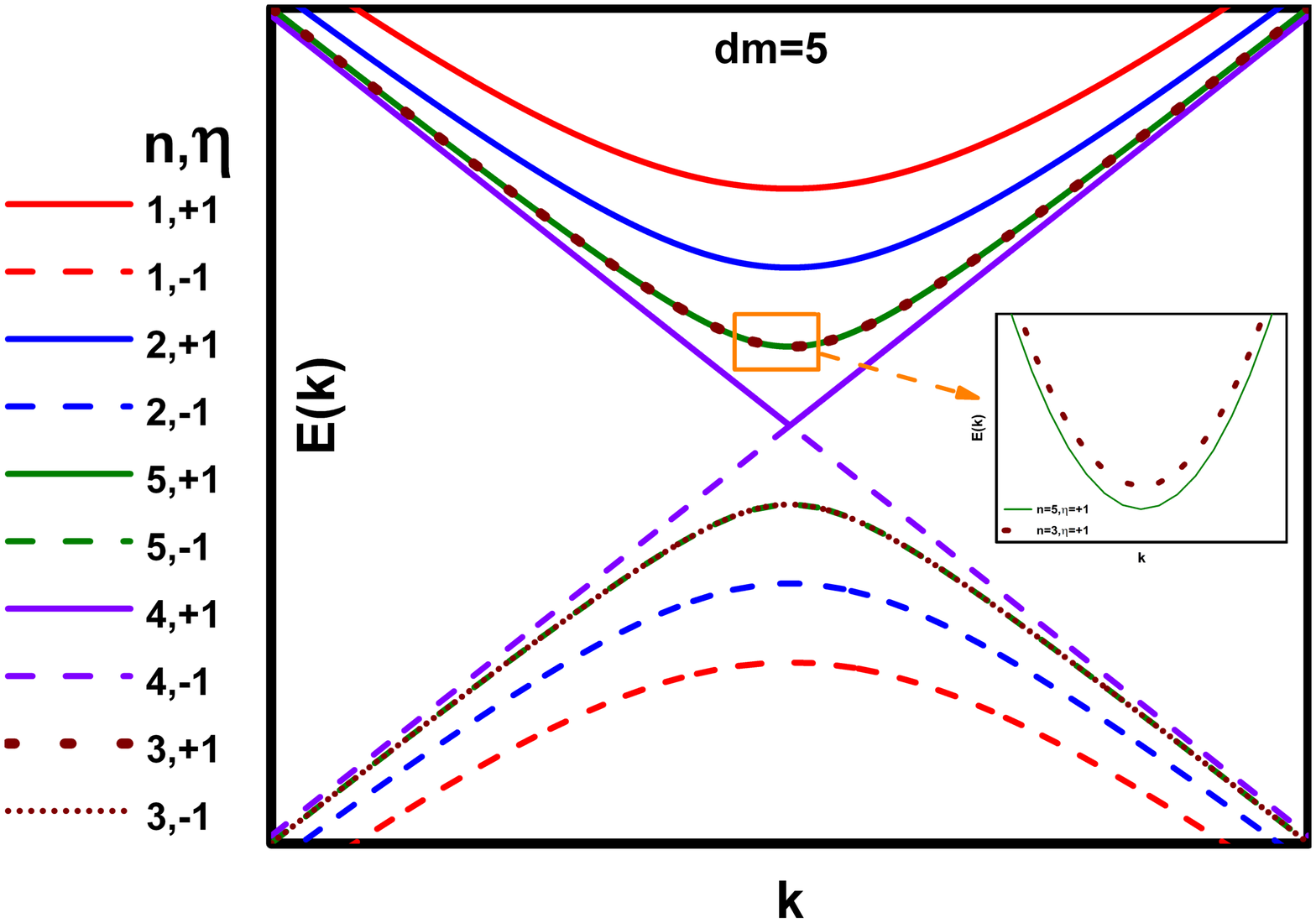}
		\vspace{-0.1cm}
		\caption{
			Single-particle energy dispersion 
			in AGNRs with  $k=a k_{x}$. The left panel is for semiconducting (dm=4) and the right one  for metallic (dm=5) AGNRs. 
			The insect  in  the right panel shows the  dispersion for $\eta=+1$ and n=3, 5 .}
		\label{d1}
	\end{figure}

	Graphene is a two-dimensional, one-atom thick planar sheet of bonded carbon atoms densely packed in a honeycomb structure as shown in the left panel of Fig. \ref{f}. In it the ribbon extends along   the $x$ axis %parallel to the armchair nanoribbon direction 
	while the graphene sheet is %quantum mechanically 
	confined along the $y$ axis. The lattice structure can be viewed as a triangular lattice with  two sites $A$ (green filled circles) and $B$ (yellow filled circles) per unit cell as shown by the rectangular box in the left panel of Fig. \ref{f}. The arrows indicate the primitive lattice vectors $\vec a_{1}= a (0, 1)$ and $\vec a_{2}= a(1/2, \sqrt{3}/2)$, with $a$ the triangular lattice constant of the structure,  and span the graphene lattice. % in 2D space. 
	Further, $\vec a_{1}$ and $\vec a_{2}$ generate the  reciprocal lattice vectors of the Brillouin zone, cf.  Fig. \ref{f}, given by $\vec b_{2}=4\pi/\sqrt{3} a  (\sqrt{3}/2, - 1/2)$ and $\vec b_{1}=4\pi/\sqrt{3} a (0, 1)$. From the explicit expressions of  $\vec b_{1}$ and $\vec b_{2}$ we find the two inequivalent Dirac points (valleys) given by $\vec K = 4 \pi/3 a(0, 1)$ and $\vec K^{\prime}=4 \pi/3 a (0, -1)$. 
	The ${\bf k\cdot p}$ Hamiltonian near the Dirac points reads %\cite{r1}
	\begin{equation}
	H = \hslash v_{F}
	\begin{pmatrix}
	0 &&  k_{-} && 0&& 0\\
	k_{+} && 0 && 0 && 0\\
	0 && 0 && 0 &&  k_{+}\\
	0 && 0 &&   k_{-} && 0
	\end{pmatrix}, \label{h1}
	\end{equation}
	%
	%%%%%%%%%%% Cite Yuya Ominato and Mikito Koshino's , R2
	where $\hslash$ is  Plank's constant, $v_{F}$ the Fermi velocity, and $k_{\pm} =  k_{y} \pm i k_{x}$. 
	%The wave function amplitude should vanish on both sublattices at the extremes, $y=0$ and $y= W$ with $W$ the width of the ribbon. \textcolor{blue}{\textcolor{red}{ the following sentence  is not correct and must be removed} To satisfy these boundary condition we must admix valleys}. 
	The resulting eigenfunctions of  Eq. (\ref{h1}) %\textcolor{red}{
	for AGNRs,  shown in Fig. \ref{ribbon}, take the form  
	\vspace*{0.2cm}
	%%%%%%%%%%%
	\begin{equation}
	\psi_{n,\eta,k_{x}}=
	\frac{1}{2 \sqrt{ L W}}
	\begin{pmatrix}
	\eta e^{-i \theta_{k_{yn},k_{x}}} e^{i k_{yn} y}\\
	e^{i k_{yn}y}\\
	%\vspace*{0.2cm}
	- \eta e^{-i \theta_{k_{yn},k_{x}}} e^{-i k_{y n}y}\\
	%\hspace{-0.5cm}%
	e^{-i k_{yn}y}
	\end{pmatrix}
	e^{ik_{x} x}\,, \quad %\quad %
	\label{h2}
	\end{equation}
	%%%%%%%%%%%  
	where $\theta_{k_{yn},k_{x}}=\tan^{-1}(k_{x}/k_{yn})$. The energy dispersion of graphene AGNRs corresponding to Eq. (\ref{h1}) is \cite{r1}
	%%%%%%%%%%%%%%%%
	\begin{eqnarray}
	E^{n}_{\eta, k_{x}}&= \eta \hslash v_{F}  \varepsilon,\,\, \,\, \varepsilon=%\sqrt{
	[k_{yn}^{2}+k_{x}^{2}]^{1/2}, % = \eta \hslash v_{F}
	\label{h3}
	\end{eqnarray}
	%%%%%%%%%%%%
	where %$\varepsilon=  \sqrt{k_{yn}^{2}+k_{x}^{2}}$, and 
	$\eta  = +1 (-1)$ stands for the conduction (valence) band. The allowed values of $k_{yn}$ are %satisfy the following condition  
	\cite{r1, r2, r3, r4}
	\begin{equation}
	k_{yn}=\dfrac{ n \pi}{W}-\dfrac{4\pi }{3 a }=\dfrac{2\pi (3n-2(dm+1))}{3a (dm+1)};
	\label{h4}
	\end{equation}
	here $W=a (dm+1)/2$ is the ribbon width,  $dm$ the number of rows of AGNRs, $a= \sqrt{3} a_{cc}$, $a_{cc}\approx 1.42 $  $\mathrm{\AA} $ is the  carbon-carbon distance, and %$n=0,\pm1,\pm2,.....$ 
	%\textcolor{red}{
	$n=1,2,...,N$ is the subband index with $N$ the  maximum number of dimmers. It follows from Eq. (\ref{h4}), if $3n-2(dm+1)=0 $, then $k_{yn}=0$ for particular $n$. So, a zero energy state appears near $k_{x}\rightarrow 0$ as in graphene, whereas the other states have band gap because $3n-2(dm+1)\neq 0 $. The energy dispersions for semiconducting $(dm=4)$ and metallic $(dm=5)$ nanoribbons are shown in Fig. \ref{d1}.
	%%%%%%%%%
	%\begin{figure}[t]
	%	\vspace*{0.5cm}
	%	\begin{center}		
	%		\includegraphics[height=3cm, width=4cm]{N-4}
	%		\includegraphics[height=3cm, width=4cm]{N-5}
	%	\end{center}	
	%	\caption{\small{(a)  Real part of the TB DDRF ($\mathfrak{Re}\chi$) for several values of $\gamma'$. (b) { 3D bar of $\mathfrak{Re}\chi$ for $\gamma'=0.05$.}  }}
	%	\label{fig :1 }
	%\end{figure}
	%%%%%%%
	
	%\subsection
	{\it Velocity matrix elements}.  To evaluate the various conductivities we need the matrix elements of the velocity operators $v_{x}=  \partial H / \hslash \partial k_{x}$ and $v_{y}=  \partial H / \hslash \partial k_{y}$.  With %The operators are %y can be readily evaluated and reads 

	\begin{equation}
	v_{x}  =
	v_{F}
	\begin{pmatrix}
	\sigma_{y} && 0\\
	0 && - \sigma_{y} 
	\end{pmatrix}, \label{v1}
	\hspace{0.1cm} 
	v_{y}= 
	v_{F}
	\begin{pmatrix}
	\sigma_{x} && 0\\
	0 && \sigma_{x} 
	\end{pmatrix},
	\end{equation}
	their matrix elements ($\left\vert  \zeta \right\rangle= \left\vert  n,\eta,k_{x} \right\rangle$) are
	%%%
	\begin{eqnarray}
	%\notag
	&&\hspace*{-0.2cm}\left\langle \zeta^{\prime}  \right\vert  v_{x} \left\vert  \zeta\right\rangle  =  N ( \eta e^{i \theta_{k_{yn}^{\prime}, k_{x}^{\prime}}} - \eta^{\prime} e^{-i \theta_{k_{yn},k_{x}}})\delta_{n,n^{\prime}} \delta_{k_{x},k_{x}^{\prime}}
	\label{v2}
	\\*
	%\end{equation}
	%%%%%%
	%%%%
	%\begin{equation}
	&&\hspace*{-0.6cm}\left\langle \zeta \right\vert  v_{y} \left\vert  \zeta^\prime \right\rangle  =  M ( \eta^{\prime}e^{i \theta_{k_{yn},k_{x}}} + \eta e^{-i \theta_{k_{yn}^{\prime},k_{x}^{\prime}}}) %(1-\delta_{n,n^{\prime}}) 
	\delta_{k_{x},k_{x}^{\prime}}, \, n\neq n^{\prime},
	\label{v3}
	\end{eqnarray}
	%%%%%
	with $e^{\pm i \theta_{k_{yn},k_{x}}} = ( k_{yn} \pm i k_{x}) / \varepsilon_{k_{x}}$, $N= - i v_{F}/2$, and $M= -i v_{F} / \pi (n-n^{\prime}) $ . %and $\left\vert  \zeta \right\rangle= \left\vert  n,s,k_{x} \right\rangle$.
	
	\subsection{Ordinary waveguides}
	
\begin{figure}[t]
		\centering
		\includegraphics[width=5cm, height=3cm%width=.87\linewidth
		]{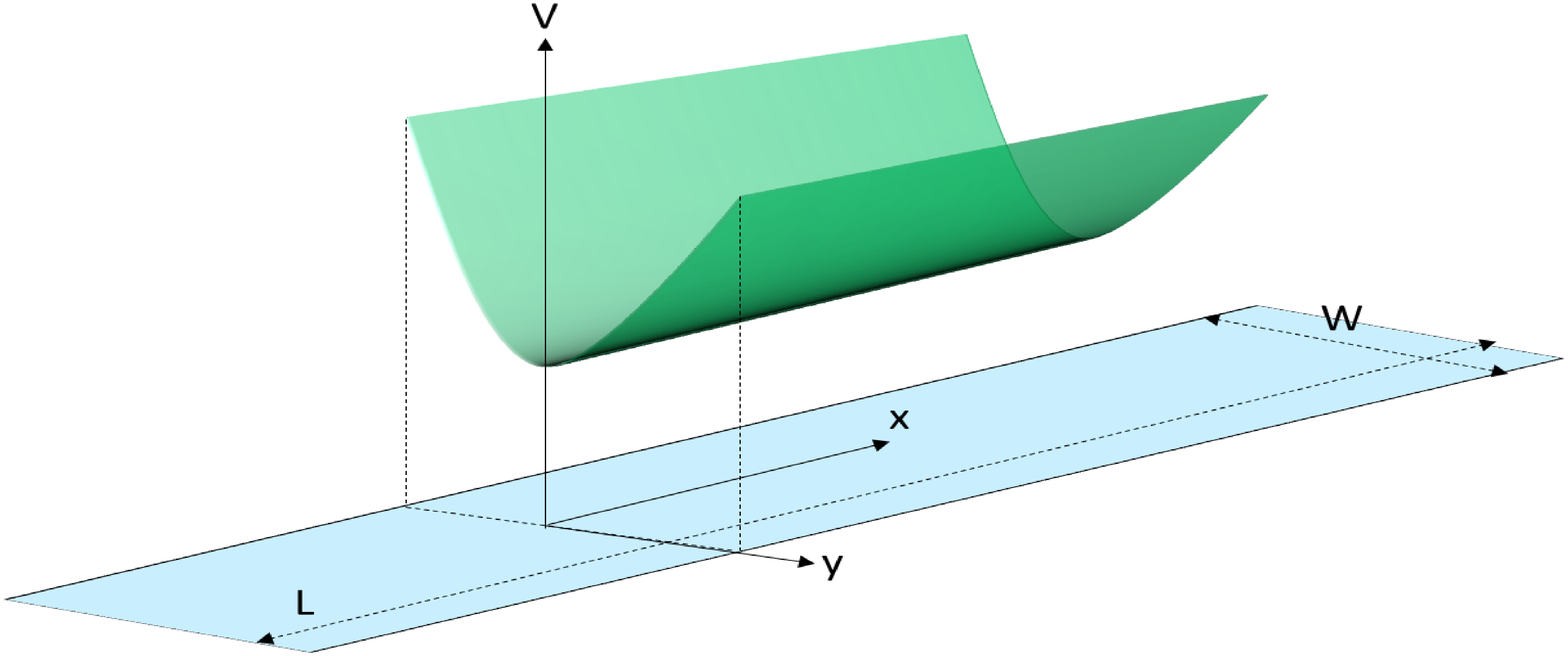}
		\hspace{0.1cm}
		%\ \\
		\includegraphics[width=3cm, height=3cm%width=.47\linewidth
		]{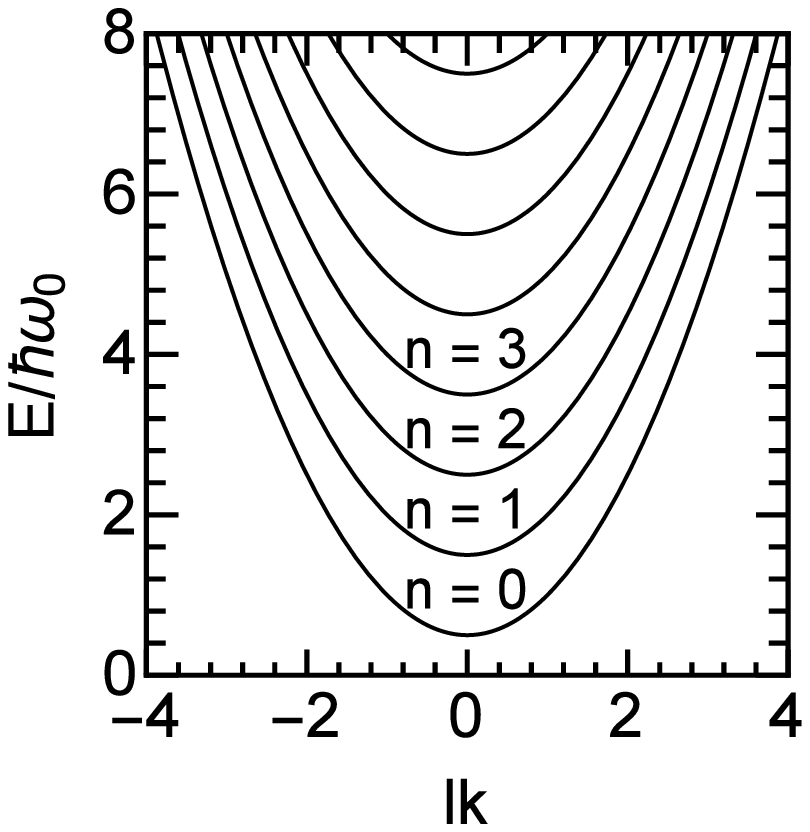}
		%\vspace{-0.1cm}
		\caption{%{\color{blue} 
			Geometry %Schematics representation 
			of a parabolically confined (left panel),  along the $y$ direction, quantum wire of width $L_y=W$ and length $L_x=L$. The right panel shows the wire's spectrum.} %with $k_{0}=\hslash /m^{\ast} \omega_{0}$ .}}% is the width. }
		\label{g1}
	\end{figure}

	In Fig. \ref{g1} we consider an ordinary %two dimensional electron gas (2DEG) 
	quantum wire along the $x$ axis %which can be 
	generated by confining a 2DEG along the $y$ direction. %Here we choose the wire plane to be the xy plane with $x$ direction parallel to the wire. 
	We assume the confining potential $V(y)$ to be %have the 
	parabolic, % form 
	i.e., $V(y)= m^{*} \omega_{0}^{2} y^{2}/2$. The  eigenvalues are %energy of ordinary waveguide in the absence of magnetic field \cite{r7} is
	%
	%%%%%%%%%%
	\begin{eqnarray}
	E_{n k_{x}}&=& (n + 1/2) \hslash \omega_{0}+%\dfrac{
	\hslash^{2} k_{x}^{2}/2 m^{*},
	\label{e1}
	\end{eqnarray}
	and the corresponding eigenfunctions
	\begin{eqnarray}
	\psi_{n k_{x}}&=& (2^{n}n ! \sqrt{\pi} \ell)^{-1/2} H_{n} (y/\ell)\, e^{-y^{2}/2 \ell^{2}} e^{i k_{x} x }
	\label{o1}
	\end{eqnarray}
	with $\ell=(\hslash /m^{\ast} \omega_{0})^{1/2}$ and $H_{n}(y/\ell)$  the Hermite polynomials. Here only the diagonal  matrix elements $v_{x}= \hslash k_{x}/m^{*}$ are relevant %here 
	since the nondiagonal ones ($\left\langle \zeta^{\prime}  \right\vert  v_{x} \left\vert  \zeta\right\rangle $) vanish. However, the nondiagonal velocity matrix elements ($\left\langle \zeta  \right\vert  v_{y} \left\vert  \zeta^{\prime} \right\rangle $) along the confinement direction are non zero and given as
	\begin{eqnarray}
	\hspace*{-0.6 cm} \left\langle \zeta \right\vert  v_{y} \left\vert  \zeta^{\prime} \right\rangle & = & N_{n} \bigl [ (n^{\prime}+1) \delta_{n^{\prime}+1,n} - (1/2) \delta_{n^{\prime}-1, n}\bigr]  \delta_{k_{x}k_{x}^{\prime}}
	\label{o2}
	\end{eqnarray}
	%{\color{blue} 
	where $N_{n}= (i \hslash/m ^{\ast} \ell  ) (2^{n^{\prime}} n^{\prime}!/2^{n} n! )^{1/2}$. It is evident from the right panel of Fig. \ref{g1} that the spectrum consists of a set of equidistant, oscillator subbands due to the %quantum 
	harmonic  confinement along the $y$ direction. %and the wave vector along the confined direction becomes discretized. 
	%Further, only semiconducting  ribbons exist in an ordinary waveguide in contrast with AGNRs.
	
	\section{ Conductivities} 
	
	We consider a many-body system described by the Hamiltonian $H = H_{0} + H_{I} - \mathbf{R \cdot F}(t)$, where $H_{0}$ is the unperturbed part, $H_{I}$ is a binary-type interaction (e.g., between electrons and impurities or phonons), and $ \mathbf{- R \cdot F}(t)$ is the interaction of the system with the external field F(t) \cite{r10}. For conductivity problems we have $\mathbf{F}(t) = e \mathbf{E}(t)$, where $\mathbf{E}(t)$ is the electric field, $e$ the electron charge, $\mathbf{R = \sum_{r_{i}}}$ , and $\mathbf{r_{i}}$  the position operator of electron $i$. In the representation in which $H_{0}$ is diagonal the many-body density operator $\rho = \rho^{d} + \rho^{nd}$ has a diagonal part $\rho^{d}$ and a nondiagonal part $\rho^{nd}$. For weak electric fields and weak scattering potentials, for which the first Born approximation applies, the conductivity tensor has a diagonal part $\sigma_{\mu\nu}^{d}$ and a nondiagonal part $\sigma_{\mu\nu}^{nd}$; the total conductivity is $\sigma_{\mu\nu}^T = \sigma_{\mu\nu}^{d} + \sigma_{\mu\nu}^{nd}, \mu,\nu = x,y$. 
	
	In general we have two kinds of currents, diffusive and hopping, with $\sigma_{\mu\nu}^{d} = \sigma_{\mu\nu}^{dif} + \sigma_{\mu\nu}^{col}$, but usually only one of them is present. If  no magnetic field is present,  the hop-\\ping  term $\sigma_{\mu\nu}^{col}$ vanishes identically  \cite{r10} and  only the term $ \sigma_{\mu\nu}^{dif} $ survives.  For  elastic scattering it is given  by % since the hopping  part $\sigma_{\mu\nu}^{col}$ vanishes identically due to the vanishing velocity matrix elements as is evident, for %quasielastic scattering, by its form
	\cite{r10, r11}
	%%%%%%%%%
	%%%%%%%
	%
	%%%%%%%%
	\begin{equation}
	\sigma_{\mu\nu}^{d} (\omega) = \dfrac{\beta e^{2}}{S_{0}} \sum_{\zeta} f_{\zeta} (1 - f_{\zeta} ) \dfrac{v_{\nu\zeta} v_{\mu\zeta} \tau_{\zeta}}{1 + i\omega \tau_{\zeta}} , \label{cc1}
	\end{equation}
	%%%%%
	where $\tau_{\zeta}$ is the momentum relaxation time, $\omega$ the frequency, and $v_{\mu\zeta}$ the diagonal matrix elements of the velocity operator. Further, $f_{\zeta} = [1 + \exp \beta (E_{\zeta} - E_{F})]^{-1}$ is the Fermi-Dirac distribution function, $\beta = 1/k_{B}T$, $T$ the temperature,  $k_{B}$ the Boltzmann constant, and $S_{0}$ the area of the sample.
	
\begin{figure}[t]
		\centering
		%\vspace*{-3cm}
		\includegraphics[width=6cm, height=3.5cm%\textwidth
		]{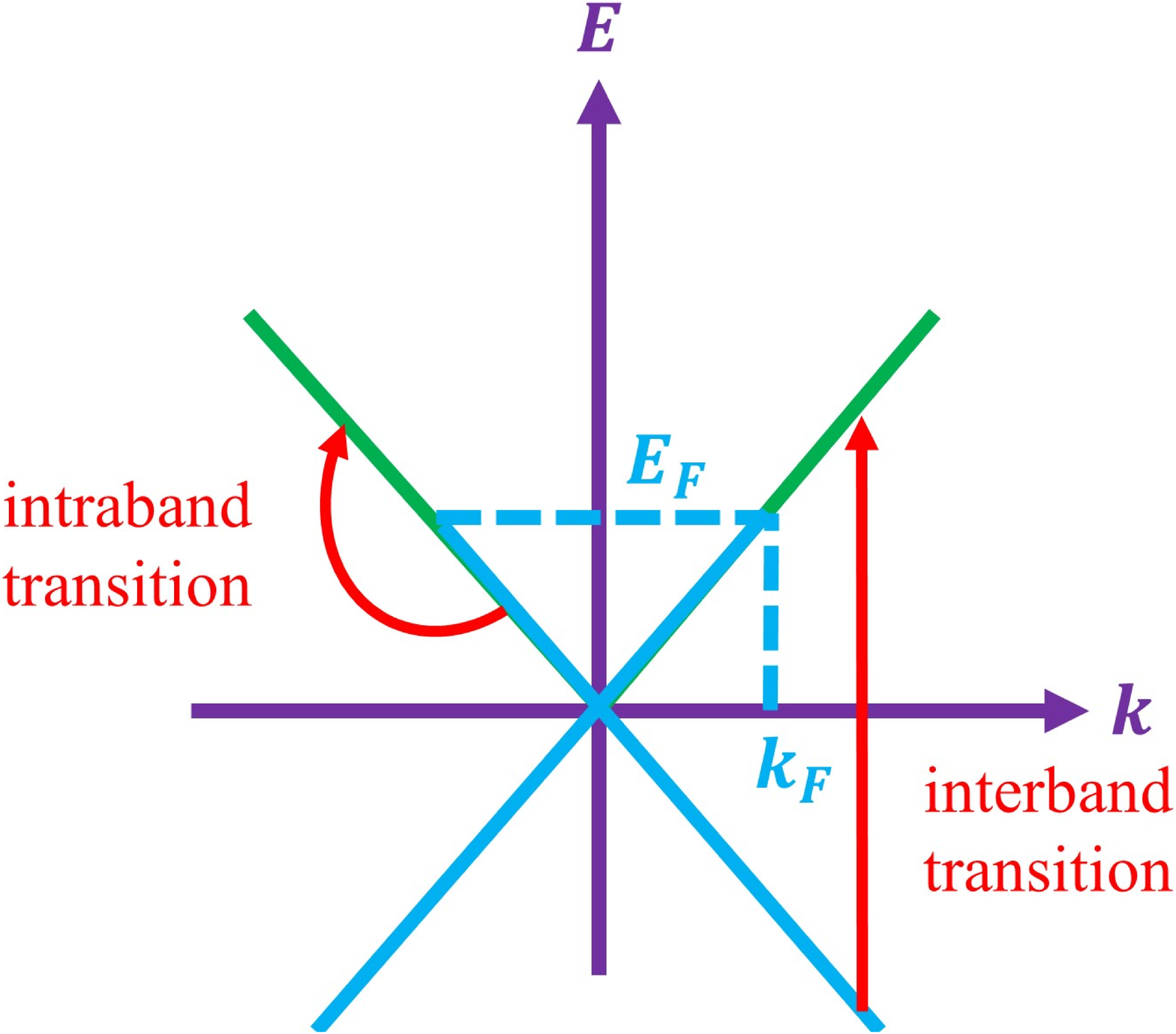}%{Fig4.eps}%{2.eps}
		%\vspace{-0.1cm}
		%\vspace*{-3cm}
		\caption{Schematic representation of  intraband and interband transitions in the energy dispersion of a metallic AGNR.}
		\label{sc}
	\end{figure}

	Regarding the contribution $\sigma_{\mu\nu}^{nd}$ one can use the identity $f_{\zeta} (1 - f_{\zeta^{\prime}})[1 - \exp \beta (E_{\zeta} - E_{\zeta^{\prime}})] = f_{\zeta} - f_{\zeta^{\prime}}$ and cast the original form in the more familiar one  \cite{r10, r11}
	%
	%%%%
	\begin{equation}
	\sigma_{\mu\nu}^{nd} (\omega) =\dfrac{ i\hslash e^{2}}{S_{0}}\sum_{\zeta \neq \zeta^{\prime}} \dfrac{(f_{\zeta} - f_{\zeta^{\prime}}) v_{\nu\zeta\zeta^{\prime}} v_{\mu\zeta\zeta^{\prime}}}{(E_{\zeta} - E_{\zeta^{\prime}})(E_{\zeta} - E_{\zeta^{\prime}} + \hslash \omega - i \Gamma )} ,\label{cc2}
	\end{equation}
	%%%%
	where the sum runs over all quantum numbers $\vert \zeta \rangle $ and $\vert \zeta^{\prime} \rangle $ with $\zeta \neq \zeta^{\prime}$. The infinitesimal quantity $\epsilon$ in the original form  \cite{r10} has been replaced by $\Gamma_{\zeta}$ to account for the broadening of the energy levels. In Eq. (\ref{cc2}) $v_{\nu \zeta \zeta^{\prime}}$ and $v_{\mu \zeta \zeta^{\prime}}$ are the nondiagonal matrix elements of the velocity operator. 
	%The total conductivity consists of diagonal and nondiagonal parts
	%
	%%%%%%%
	%\begin{equation}
	%\sigma_{\mu\nu}^{t}= \sigma_{\mu\nu}^{d}+\sigma_{\mu\nu}^{nd}, \label{v4}
	%\end{equation}
	%%%%%%%
	Further, diagonal and nondiagonal contributions describe  intraband and interband  %electron 
	transitions, respectively, as shown %. These transitions are 
	schematically in Fig. \ref{sc}.
	%%%%%%%%%%%
	%%%%%%%
	%
	%%%%%%%%%%%%
	\begin{figure}[t]
		%\centering
		\includegraphics[width=8cm, height=5cm%width=.87\linewidth
		]{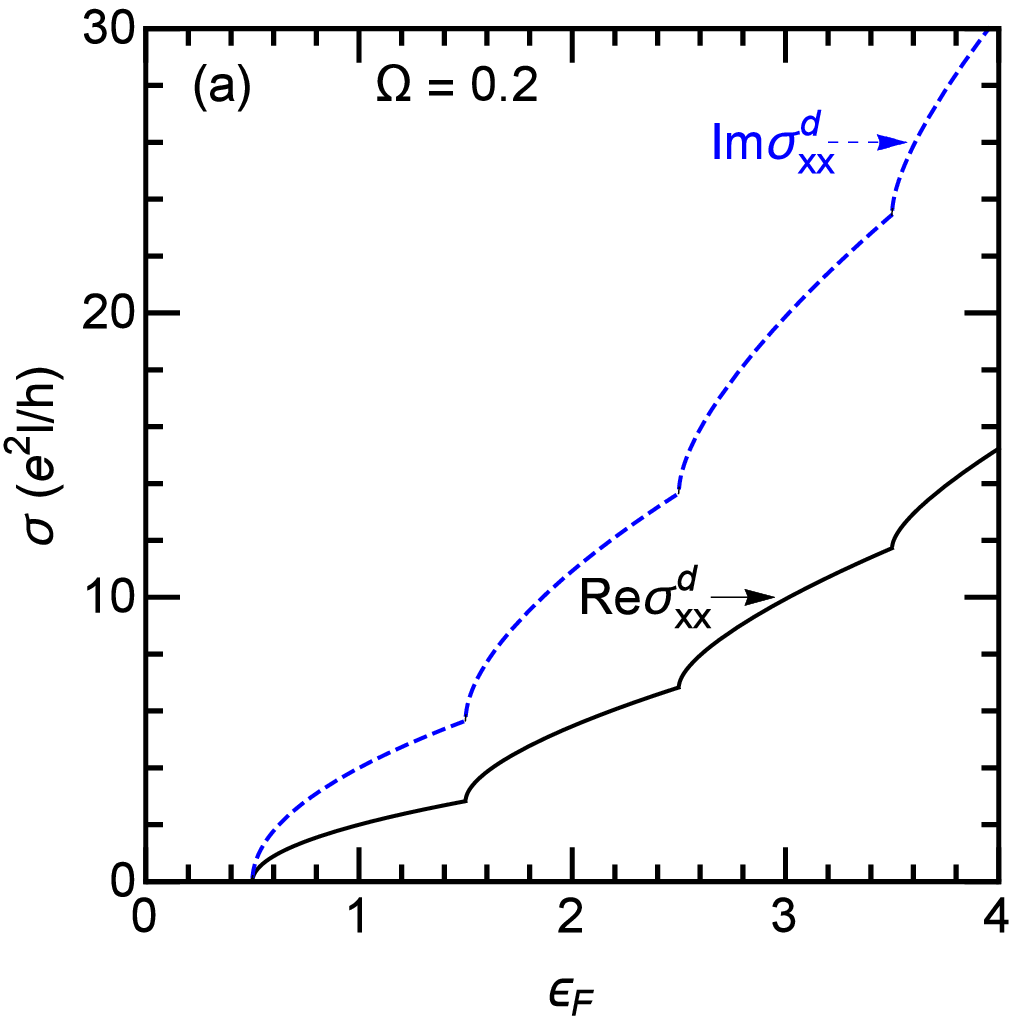}
		\ \\
		\ \\
		\includegraphics[width=8cm, height=5cm%width=.87\linewidth
		]{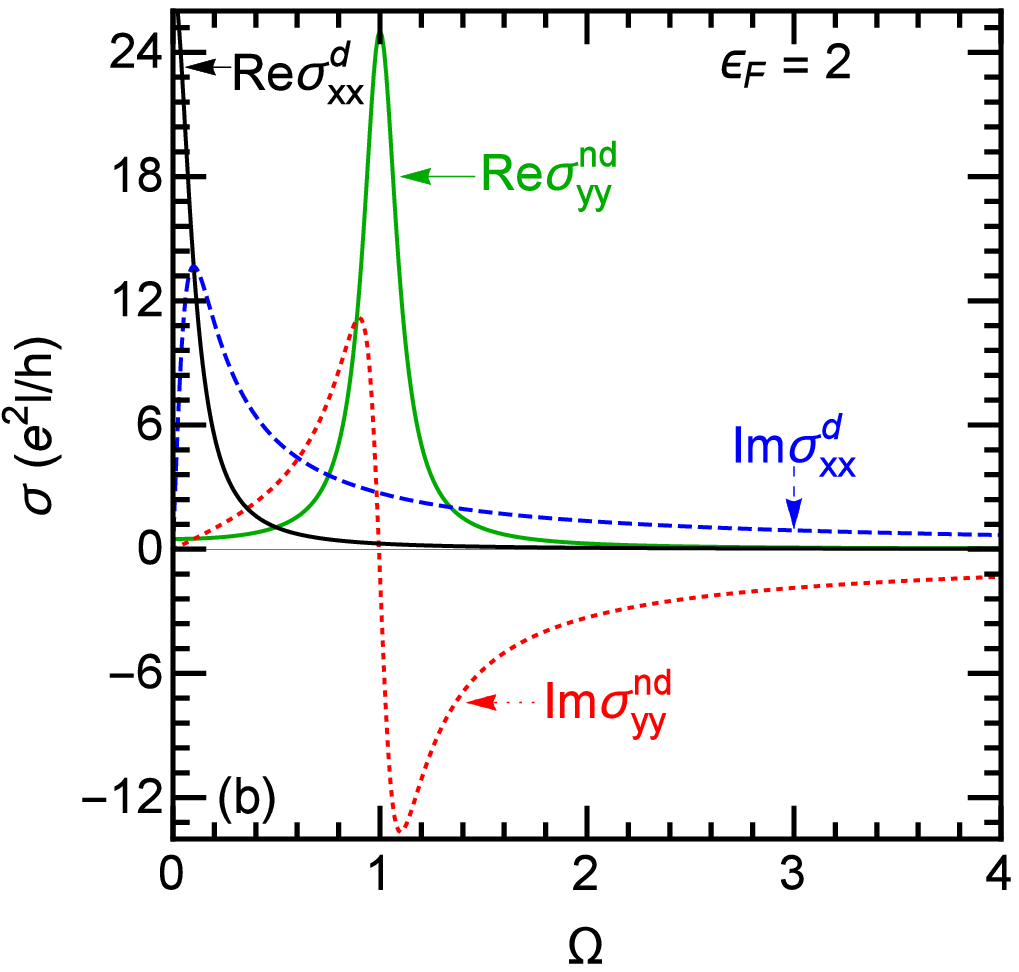}
		%\vspace*{1cm}
		\vspace{-0.2cm}
		\caption{ Diagonal conductivity, in units of  %$\sigma_{0}=e^{2}/\pi \hslash \sqrt{2 m^{\ast}}$, 
			$e^{2} \ell /h$ of an ordinary waveguide vs $\varepsilon_{F}=E_{F}/\hslash \omega_{0}$ in (a) and  vs photon energy $(\Omega=\omega/\omega_{0})$  in (b). The black (blue) curves are for $\mathrm{Re}\sigma_{xx}^{d}$ ($\mathrm{Im}\sigma_{xx}^{d}$) and the dark green (red dotted) ones are for $\mathrm{Re}\sigma_{yy}^{nd}$ ($\mathrm{Im}\sigma_{yy}^{nd}$). Here we used $\gamma=\Gamma/ \hslash \omega_{0}=0.1$. %Here we used 
			%$\sigma_{0}=e^{2}/\pi \hslash \sqrt{2 m^{\ast}}$.  
		}\label{ord}
	\end{figure}
	%%%%%%%%%%	
	\subsection{Diagonal conductivity in ordinary  waveguides}
	
	For $\omega=0$ %, $\vert \zeta  \rangle = \vert n,k_{x} \rangle$ 
	and $\mu=\nu=x$  Eq. (\ref{cc1})  becomes
	%
	%%%%%%%%%%%
	\begin{eqnarray}
	\sigma_{xx}&=& \dfrac{ \beta e^{2}}{L} \sum_{n k_{x}} f_{n k_{x}}(1-f_{n k_{x}}) \, v_{x}^{2} \,\tau_{n k_{x}}.
	\label{e2}
	\end{eqnarray}
	For very low temperatures, we  make the approximation $\beta f_{n k_{x}}(1-f_{nk_{x}})\approx\delta(E_{n k_{x}}-E_{F})$, replace $\tau_{\zeta}$ by $\tau_{F}$,  and use the prescription $\sum_{k_{x}}\rightarrow (L_{x}/2 \pi) \int dk_{x}$. Then  Eq. (\ref{e2}), with  $v_{x}= \hslash k_{x}/m^{*}$, takes the form 
	\begin{eqnarray}
	\sigma_{xx}^{d}(i \omega) & = & \dfrac{\sigma_{0}  \tau_{F} }{(1+i \omega \tau_{F})}\sum_{n} \sqrt{E_{F}-E_{n}}, 
	\label{e5}
	\end{eqnarray}
	%%%%%%%%%%%  
	where $E_{n}=(n+ 1/2) \hslash \omega_{0}$ and $\sigma_{0}=e^{2}/\pi \hslash \sqrt{2 m^{\ast}}$.  For the dc conductivity we simply set $\omega=0$.
	
	In Fig. \ref{ord} we  show the diagonal conductivity as a function of $E_{F}$ (upper panel) and photon energy (lower panel) for $\hslash \omega_{0}=0.5$ meV \cite{nr11}. The conductivity increases with the increase of $E_{F}$ but cusps appear due to the presence of discrete levels in the lateral direction produced by the parabolic confinement. In addition, $\sigma_{xx}^{d}$ vanishes when the Fermi level is in the range $0 \leqslant \varepsilon_{F}\leqslant 0.5$ %\textcolor{blue}{
	since the electron density is null in this range of energy. We can see that $\mathrm{Re}\sigma_{xx}^{d}$ has a Drude-type peak around $ \Omega=0$ while $\mathrm{Im}\sigma_{xx}^{d}$ has peak around $ \Omega= 0.1$ as can be seen in the lower panel of Fig. \ref{ord}.  Furthermore, it can also be seen that the Drude-type contribution survives at low frequencies while it vanishes at higher frequencies. Note that the nondiagonal contribution %{\color{blue}$\sigma_{xx}^{nd}$}
	$\sigma_{xx}^{nd}$ to the conductivity of 2DEG when confined in a ribbon vanishes,  since the velocity matrix elements are diagonal,  whereas we will find below that it survives in graphene ribbons.
	
	\subsection{Nondiagonal conductivity in ordinary  waveguides}
	
	With the help of  matrix elements (\ref{o2}) and % $  \left\vert  \zeta^{\prime} \right\rangle=  \left\vert  n, k_{x} \right\rangle$,
	$  \left\vert  \zeta \right\rangle=  \left\vert n, k_{x} \right\rangle$, we can recast  Eq. (\ref{cc2}) as
	%%%%%%%%%
	\begin{eqnarray}
	\notag
	\hspace*{-0.6cm}\sigma_{yy}^{nd}(i \omega) & = &- \dfrac{i  e^{2}}{4 \pi \sqrt{2 m^{\ast}}} \sum_{n} \int_{\vert E_{n} \vert}^{E_{m}} d E \dfrac{(n+1)(f_{k_{x}}^{n}-f_{k_{x}}^{n+1})}{[E-E_{n}]^{1/2}}
	\notag
	\\ & &
	\times \biggl[ \dfrac{\hslash \omega_{0}+ \hslash \omega + i \Gamma}{(\hslash \omega_{0} + \hslash \omega)^{2} + \Gamma^{2}}  -  \dfrac{\hslash \omega_{0} - \hslash \omega -  i \Gamma}{(\hslash \omega_{0} -  \hslash \omega)^{2} + \Gamma^{2}}\biggr]
	\label{o3}
	\end{eqnarray}
	%%%%%%%%
	where $E_{m}=E_{n}+ \hslash^{2} k_{m}^{2}/2 m^{\ast}$. In the limit $\Gamma=\omega=0$, one can show that $\sigma_{yy}^{nd}(i \omega)$   vanishes.
	
	In Fig. \ref{ord} (b), we have plotted the numerically evaluated $\mathrm{Re}\sigma_{yy}^{nd}$ (dark green curve) and $\mathrm{Im}\sigma_{yy}^{nd}$ (red dotted curve)  as  functions of the dimensionless photon energy $(\Omega= \omega/ \omega_{0})$. We can see that $\mathrm{Re}\sigma_{yy}^{nd}$ is finite at $\Omega=0$, due to $\Gamma\neq 0$, and attains a maximum value at $ \Omega= 1$. Upon further increasing $\Omega (\geqslant 1)$ we see that  $\mathrm{Re}\sigma_{yy}^{nd}$ approaches to zero. On the other hand, we observe that $\mathrm{Im}\sigma_{yy}^{nd}$ acquires positive and negative values  due to the $\hslash \omega_{0}-\hslash \omega$ factor in Eq. (\ref{o3}). For $\hslash \omega_{0} > \hslash \omega$, the second term of Eq. (\ref{o3}) is greater than the first one and we find the positive peak. However, we obtain a negative absorption peak for  $\hslash \omega_{0} < \hslash \omega$. It can also be seen from Eq. (\ref{o3}) that only intraband  transitions occur in contrast to AGNRs where both intraband and interband transitions occur, see  Eqs. (\ref{v17})-(\ref{v18}) below.
	
	\subsection{Diagonal conductivity in AGNRs } 
	
	%\subsubsection{ 
	{\bf $\tau$ constant}.
	From Eq. (\ref{h3}) we readily find the velocity%component of the 
	% as %matrix element from Eq. (\ref{h3}) is %calculated as $(v_{x}= \partial E_{\eta k_{x}}^{n} / \hslash \partial k_{x})$
	%
	\begin{equation}
	v_{x}=\eta  v_{F} k_{x}/\varepsilon. \label{v5}
	\end{equation}
	
	Substituting  Eq. (\ref{v5})  in  Eq. (\ref{cc1}), using $\beta f_{\eta  k_{x}}^{n}(1-f_{\eta k_{x}}^{n})\approx\delta(E_{\eta  k_{x}}^{n}-E_{F})$ and $\tau_{\eta k_{x}}^{n}=\tau_{F}$ at zero temperature, and performing the integration over $k_{x} ((L_{x}/2 \pi) \int dk_{x})$, we find the  conductivity expression of AGNRs for %zero and 
	finite $\omega$ as 
	\begin{eqnarray}
	\sigma_{xx}^{d}(i \omega) =  \dfrac{ e^{2} v_{F}\tau_{F}}{h(1+i \omega \tau_{F})} \sum_{n}\,%\dfrac{
	%\sqrt{ 
	\frac{X_{Fn }%[\varepsilon_{F}^{2}-  k_{yn}^2]^{1/2}\big
	}{ \varepsilon_{F}},
	\label{v7}
	\end{eqnarray}
	%%%%%%%
	where $X_{Fn}= [\varepsilon_{F}^{2}-  k_{yn}^2]^{1/2}, \varepsilon_{F}=E_{F}/\hslash v_{F}$, and the summation terminates at the last occupied level. %In addition, 
	Equation (\ref{v7}) is only valid for $\varepsilon_{F} \geqslant k_{yn}$. For $k_{yn}=0$  it %Eq. (\ref{v7}) %-(\ref{v7})  
	reduces to
	%%%%%
	\begin{equation}
	%\sigma_{xx}^{d} (0)= \dfrac{ e^{2} v_{F} \tau_{F}}{h}  n_{F},  \quad
	\sigma_{xx}^{d}(i \omega) =  \dfrac{ e^{2} v_{F}\tau_{F}}{h(1+i \omega \tau_{F})}\, n_{F},
	\label{v8}
	\end{equation}
	%%%%%%%
	where $n_{F}$ is the number of occupied levels.
	\begin{figure}[t]
		\centering
		\includegraphics[width=9cm, height=6cm%width=.45\textwidth
		]{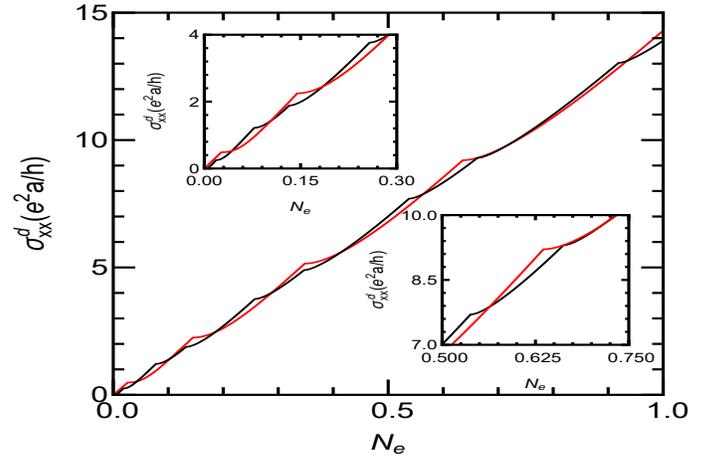}
		\vspace{-0.2cm}
		\caption{Conductivity $\sigma_{xx}^{d}$ for screened Coulomb scatterers as a function of the dimensionless carrier density $(N_{e}=an_{e}/2\pi)$ for semiconducting (black curves) and metallic (red curves) nanoribbons. Cusps in the curves  appear  when new subbands are occupied by increasing %electrons at critical values of 
			the electron density. For further clarity the ranges $0-0.3$  and $0.5-75$ are shown in the insets.}
		\label{dc1}
	\end{figure}
	{\bf $ \tau \neq$ constant}.
	{\it Long-range impurities}. Using  Eqs. (\ref{v5}),  (\ref{t4}), and the same assumptions, as given above Eq. (\ref{v7}), in Eq. (\ref{cc1}) we obtain for $\omega=0$ and  $\eta=+1$
	%
	%%%%%%%%
	\begin{equation}
	\sigma_{xx}^{d} (0)  =  \dfrac{ e^{2} A}{h} \sum_{ n}    \dfrac{ X_{Fn}^2%\varepsilon_{ F}^{2}-k_{yn}^{2}
	}{ \varepsilon_{ F}^{2}} \Big[\dfrac{k_{s}^{2}+4X_{Fn}^2%( \varepsilon_{ F}^{2}-k_{yn}^{2})
	}{k_{s}+%\sqrt{
		[k_{s}^{2}+4X_{Fn}^2%( \varepsilon_{ F}^{2}-k_{yn}^{2})
		]^{1/2}} \Big] ,\label{v9}
	\end{equation}
	%%%%%%%%%
	with $A=   2  \hslash^2 v_{F}^{2}/ n_{i}U_{s}^{2} $. % and $X_{Fn}^2=\epsilon_{ F}^{2}-k_{yn}^{2}$. 
	For $k_{yn}=0$  Eq. (\ref{v9}) becomes 
	%
	%%%%%%%%
	\begin{align}
	\sigma_{xx}^{d} (0)
	= \dfrac{ e^{2} A}{h } \dfrac{k_{s}^{2}+4 \varepsilon_{ F}^{2}}{k_{s}+%\sqrt{
		[k_{s}^{2}+4 \varepsilon_{ F}^{2}]^{1/2}} \,n_{F}. \label{v10}
	\end{align}	
	%\subsubsection{ 
	{\it Short-range impurities}.  We consider the potential $U(x)=U_{0}\,\delta(x-x_{i})$ with $U_{0}$ its constant strength  and $x_{i}$ the position of the impurity.
	Corresponding to Eq. (\ref{v9}) we find the dc conductivity is now given by %
	%%%%%%%%
	\begin{eqnarray}
	%\vspace*{-2cm}
	\sigma_{xx}^{d} (0) & = & %\dfrac{ 
	\dfrac{ e^{2} B}{h } %(e^{2}/h) B 
	\sum_{ n} \,\,  % \dfrac{ 
	\frac{X_{Fn}^2%\varepsilon_{ F}^{2}-k_{yn}^{2}]
	}{ \varepsilon_{ F}^{2}} ,\label{v11}
	\end{eqnarray}
	%%%%%%%%%
	where $B=\pi  \hslash^2 v_{F}^{2} /  2 n_{i}U_{0}^{2}$. For $k_{yn}=0$ Eq. (\ref{v11}) becomes  
%
	%%%%%%%%
	\begin{align}
	&\sigma_{xx}^{d} (0)
	=%(e^{2}/h) B % 
	\dfrac{ e^{2} B}{h } 
	n_{F}. \label{v12}
	\end{align}
	%%%%%%%%%
	
	For the finite frequency $\omega$ results we simply divide those of Eqs. (\ref{v9})-(\ref{v12}) by $1+i\omega\tau_F$.
	\begin{figure}[t]
		\centering
		\includegraphics[width=3.5cm, height=3cm%width=.48\linewidth
		]{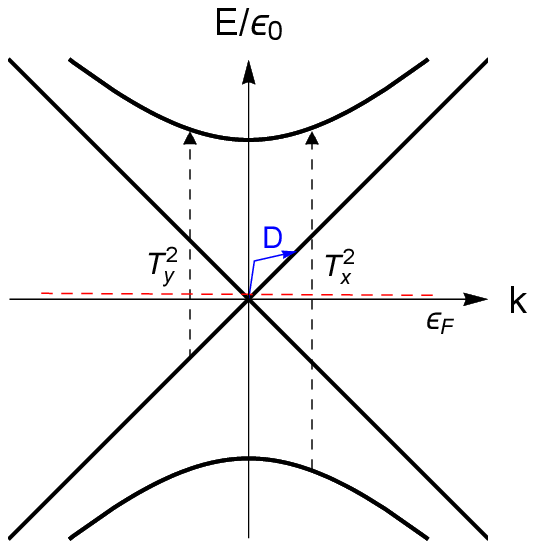}
		\hspace*{0.3cm}
		\includegraphics[width=3.5cm, height=3cm%width=.48\linewidth
		]{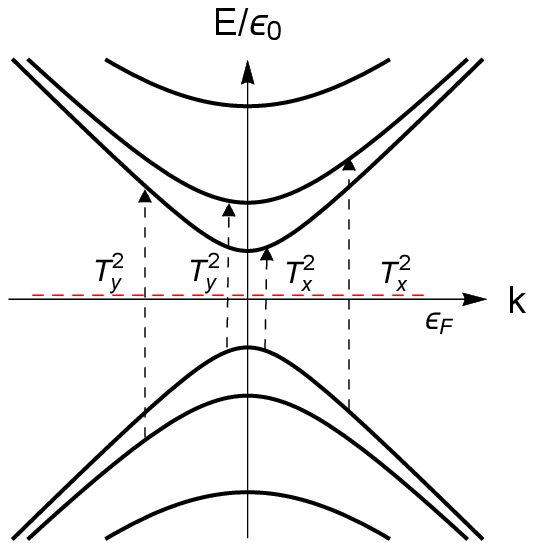}
		\ \\
		\ \\
		\includegraphics[width=3.5cm, height=3.5cm%width=.48\linewidth
		]{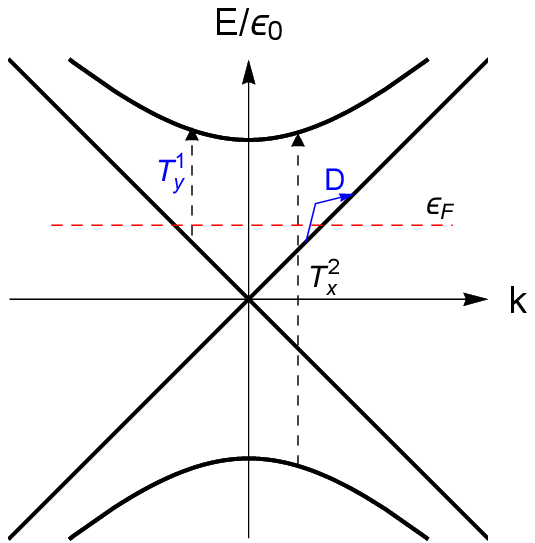}
		\hspace*{0.3cm}
		\includegraphics[width=3.5cm, height=3.5cm%width=.48\linewidth
		]{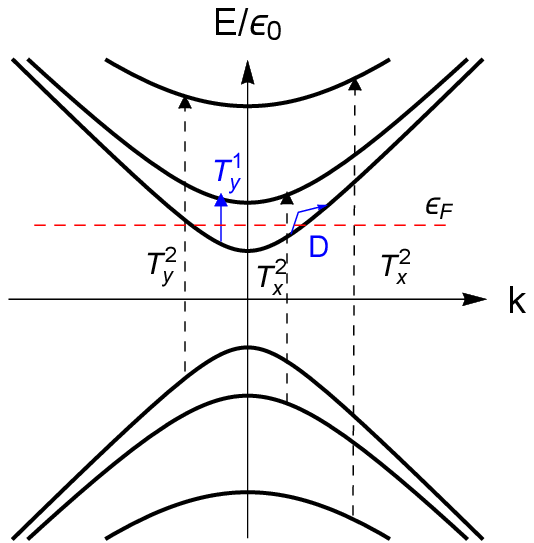}
		%\vspace{-0.1cm}
		\caption{Schematic representation of some allowed %optical 
			transitions indicated by arrows. The horizontal red dashed lines show  the Fermi level.  $T_{y}^i$ and $T_{x}^i$ denote intraband  $(i=1)$ and interband $(i=2)$ transitions, respectively at the peaks of  $\sigma_{yy}$ and $\sigma_{xx}$ in Fig. \ref{nd1}, see Eqs. (\ref{v15})-(\ref{v18}), while $D$ represents the Drude-type intraband transition, cf. Eq. (\ref{v7}).}
		\label{tt1}
	\end{figure}
	%%%%%%%
	
	In Fig. \ref{dc1}, we plot $\sigma_{xx}^{d}$ as a function of the dimensionless carrier density $(N_{e}=a n_{e}/2\pi)$ for $dm=4$ (black line) and $dm=5$ (red line). The relevant relaxation time is given by Eq. (\ref{t4}) in appendix. The %We may identify the screening 
	factor $k_{s}$ can be approximated by the Thomas-Fermi wave vector $k_{s}= (2\pi e^{2}/\epsilon) D(E_{F})$ with $\epsilon$  the relative dielectric constant and $D(E_{F})$ the density of states at the Fermi level. We can see that $\sigma_{xx}^{d}$ increases almost linearly from $0$ with the peaks at critical value of $N_{e}$. These peaks appear when the subbands start to be occupied by electrons,.
	%whereas it tends to increase linearly in the gap between the subbands. 
	Also, this behaviour is consistent with the band structures, cf. Fig. \ref{d1}. These jumps are absent in the conductivity of graphene \cite{r12}. Further, we observe the richer structure of peaks for semiconducting ribbons than metallic ones due to the opening of gaps among the subbands of semiconducting nanoribbons as can be seen by comparing the left and right panels of Fig. \ref{d1}.   It is worth mentioning that this scattering-dependent contribution was not accounted for in previous studies, see, e.g., Refs. \cite{r2}, \cite{r4}.
	%%%%%%%%%%%%%%
%\vspace*{-1cm}
	\subsection{Nondiagonal conductivity in AGNRs} 
	With $\vert \zeta  \rangle=\vert n, \eta, k_{x}  \rangle $ Eq. (\ref{cc2}) %and  $\vert \zeta ^{\prime} \rangle=\vert n^{\prime}, \eta^{\prime},k_{x}^{\prime}  \rangle $ 
	becomes
	%
	%%%%%%%
	\begin{eqnarray}
	\notag
	\sigma_{xx}^{nd}(i\omega) & = & \dfrac{i\hslash e^{2}}{L_{x}}  \sum_{\eta \eta^{\prime} n n^{\prime} k_{x} k_{x}^{\prime}} \dfrac{f_{\eta k_{x}}^{n}-f_{\eta^{\prime} k_{x}}^{n^{\prime}} }{E_{\eta k_{x}}^{n}-E_{\eta^{\prime} k_{x}}^{n^{\prime}} } 
	\\ & &  \times \dfrac{ v_{x \eta \eta^{\prime} k_{x}}^{n n^{\prime}} v_{x  \eta^{\prime} \eta k_{x}}^{n^{\prime}n}}{ E_{\eta k_{x}}^{n}-E_{\eta^{\prime} k_{x}}^{n^{\prime}}+ \hslash \omega+ i \Gamma_{\eta \eta^{\prime} k_{x} k_{x}^{\prime}}^{n n^{\prime}} },\label{v13}
	\end{eqnarray}
	%%%%%%%%%%%%
	where  $v_{x \eta \eta^{\prime} k_{x}}^{n n^{\prime}}=\langle n^{\prime}, \eta^{\prime},k_{x}^{\prime}\vert  v_{x} \vert n, \eta, k_{x} \rangle$ and $v_{x  \eta^{\prime} \eta k_{x}}^{n^{\prime}n}=\langle n^{\prime}, \eta^{\prime}, k_{x}^{\prime}\vert  v_{x} \vert n, \eta, k_{x} \rangle$  are the nondiagonal matrix elements of the velocity operator. Further, the velocity matrix element (\ref{v2}) is diagonal in $k_{x}$, therefore $k_{x}$ will be suppressed in order to simplify the notation. The summation in Eq. (\ref{v13}) runs over all quantum numbers $n$,$n^{\prime}$, $\eta$, $\eta^{\prime}$, and $k_{x}$. The parameter $\Gamma_{\eta \eta^{\prime}}^{n n^{\prime}}$, that takes into account the level broadening, is assumed to be independent of the band and subband indices i.e. $\Gamma_{\eta \eta^{\prime}}^{n n^{\prime}}=\Gamma$. Also, we will simplify the notation over summation by considering the subband orthogonality $\delta_{k_{yn}k_{yn}^{\prime}}$. Hence,  after expanding the fraction,   Eq. (\ref{v13}) can be rewritten as
	
	%%%%%%%
	\begin{eqnarray}
	\notag
	\sigma_{xx}^{nd}(i\omega) & = & \dfrac{i\hslash e^{2}}{L_{x}}  \sum_{\eta \eta^{\prime} n k_{x}} \dfrac{(f_{\eta k_{x}}^{n}-f_{\eta^{\prime} k_{x}}^{n})  v_{x \eta \eta^{\prime} k_{x}}^{n n} v_{x  \eta^{\prime} \eta k_{x}}^{n n}}{E_{\eta k_{x}}^{n}-E_{\eta^{\prime} k_{x}}^{n}} 
	\\ & &
	\times \dfrac{ E_{\eta k_{x}}^{n}-E_{\eta^{\prime} k_{x}}^{n}+ \hslash \omega - i \Gamma}{  (E_{\eta k_{x}}^{n}-E_{\eta^{\prime} k_{x}}^{n}+ \hslash \omega)^{2}+  \Gamma^{2} }\,. \label{v14}
	\end{eqnarray}
	%%%%%%%%%%%%
	\begin{figure}[t]
		\centering
		\includegraphics[width=8cm, height=5cm%width=.87\linewidth
		]{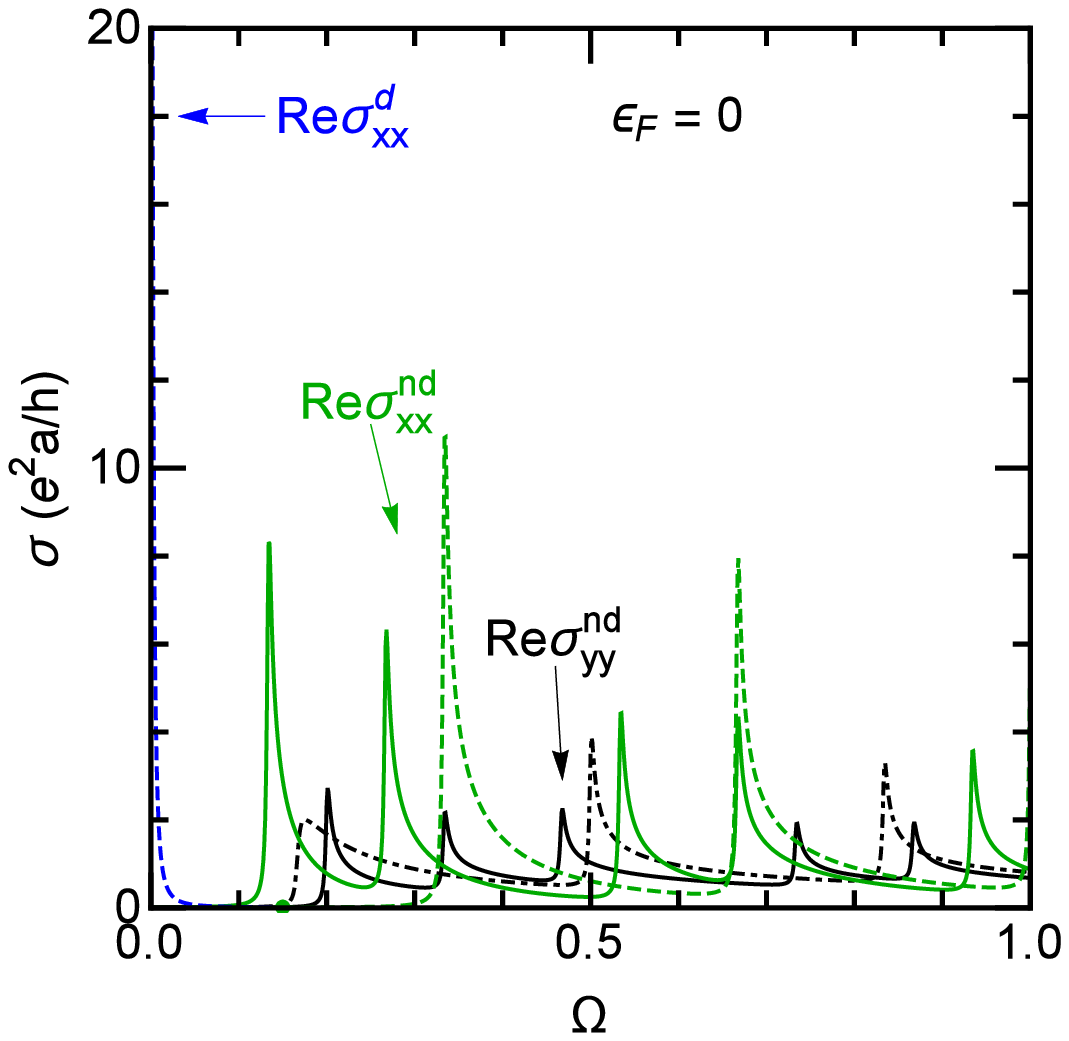}
		\ \\
		\ \\
		\includegraphics[width=8cm, height=5cm%width=.87\linewidth
		]{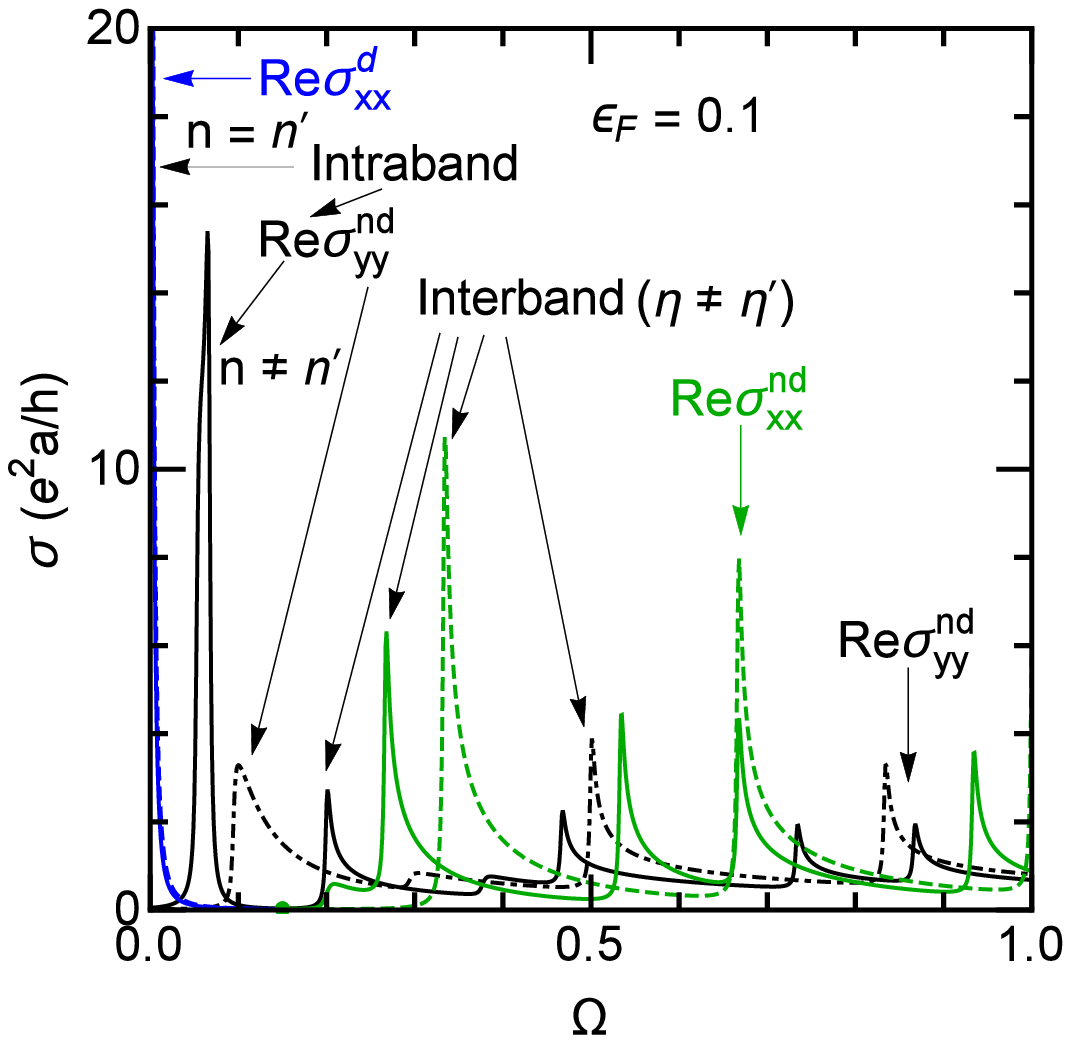}
		\vspace{-0.1cm}
		\caption{Real part of the conductivity vs frequency %photon energy 
			for $\varepsilon_{F} = 0$ (upper panel) and   $\varepsilon_{F} = 0.1$ (lower panel,  $k_{B} T/\varepsilon_{0}=0.001$, and $ \Gamma/\varepsilon_{0}=0.002$. The solid curves are for semiconducting nanoribbons ($dm=4$) and the dotted ones are for metallic ribbons ($dm=5$).} %The peak positions follow the % same 
		%order  indicated in Fig. \ref{tt1}.  }
		\label{nd1}
	\end{figure}
	%%%%%%%
	
	We  evaluate  Eq. (\ref{v14}) by considering the summation over $\eta=+1$, $\eta^{\prime}=-1$, and $\eta=-1$, $\eta^{\prime}=+1$, denoted by $\sum_{-+}$ and $\sum_{+-}$.  For $\eta = \eta^{\prime}$ the contributions $\sum_{++}$ and $\sum_{--}$ to $\mathrm{Re} \sigma_{xx}^{nd}(i\omega)$ are not allowed due to the condition $\zeta\neq\zeta^\prime$, cf. Eqs. (6), (11). Hence, the summation over $\eta= \eta^{\prime}$ is given only by the Drude-type, intraband contribution $\sigma_{xx}^d( i\omega)$ to the total conductivity, see Eqs. (15), (17). %Therefore, we have treated it ($\sum_{++}$ and $\sum_{--}$) in subsection $\bf{B}$ separately.  So, t
	
	The real and imaginary parts corresponding to Eq. (\ref{v14}) read
	%%%
	\begin{eqnarray}
	\notag
	\mathrm{Re}\sigma_{xx}^{nd}(i\omega) & = & - \dfrac{ e^{2} v_{F} }{4 \pi} \sum_{n} \int_{\vert  k_{yn}\vert}^{\varepsilon_{m}} d\varepsilon \dfrac{ k_{yn}^{2} (f_{- k_{x}}^{n}-f_{+ k_{x}}^{n})}{\varepsilon^{2} 
		%\sqrt{\varepsilon^{2}-  k_{yn}^{2} } }
		[\varepsilon^{2}-  k_{yn}^{2}]^{1/2}}
	\\ & &
	\times  (C_{+} + C_{-}), \label{v15}
	\end{eqnarray}
	%%%%%
	and
	%%%
	\begin{eqnarray}
	\notag
	\mathrm{Im}\sigma_{xx}^{nd}(i\omega) & = & - \dfrac{ e^{2} v_{F} }{4 \pi} \sum_{n} \int_{\vert  k_{yn}\vert}^{\varepsilon_{m}} d\varepsilon \dfrac{ k_{yn}^{2} (f_{- k_{x}}^{n}-f_{+ k_{x}}^{n})}{\varepsilon^{2} 
		%\sqrt{
		[\varepsilon^{2}-  k_{yn}^{2}]^{1/2} }
	\\ & &
	\times (R_{+} - R_{-}), \label{v16}
	\end{eqnarray}
	%%%%%
	%%%%%%%%%%%%
	\begin{figure}[t]
		\centering
		\includegraphics[width=8cm, height=5cm%width=.87\linewidth
		]{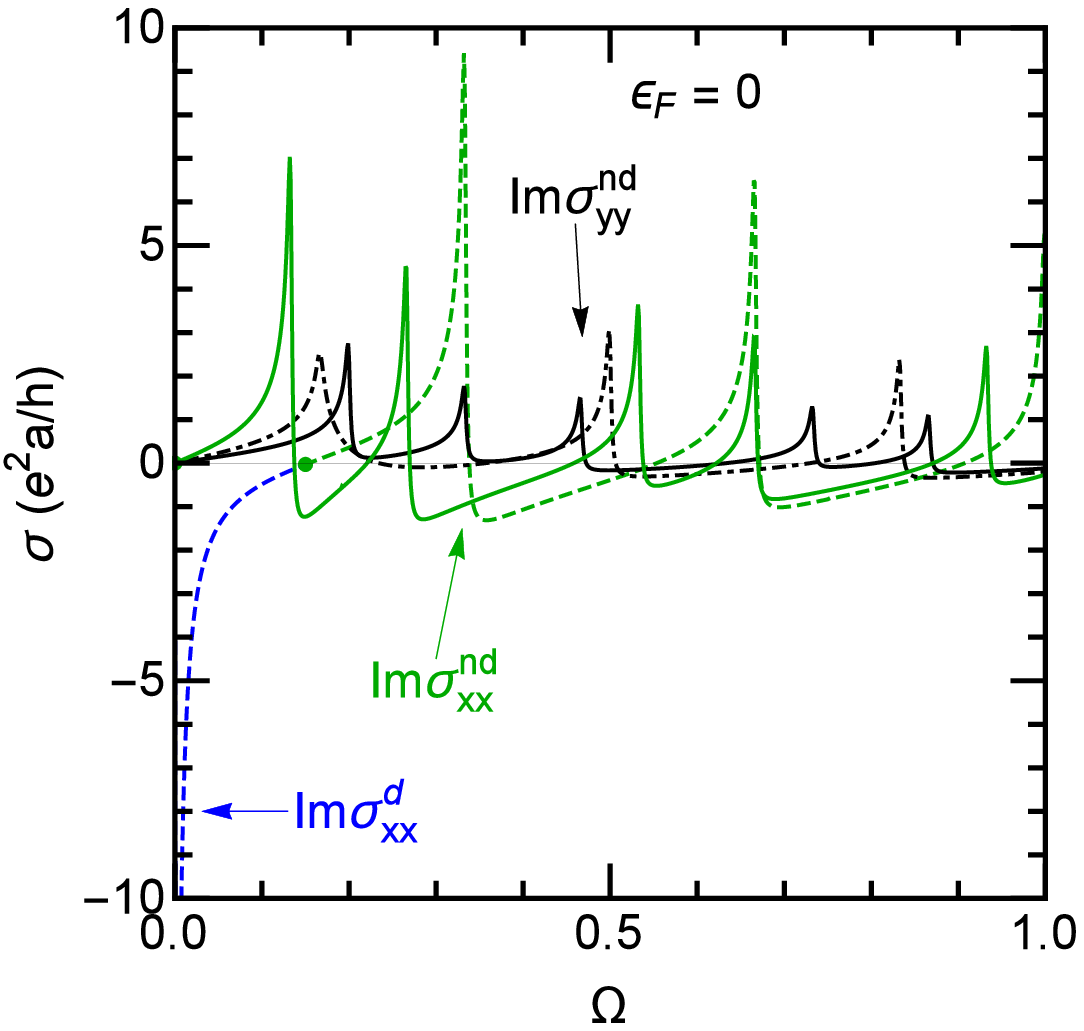}
		\ \\
		\ \\
		\includegraphics[width=8cm, height=5cm%width=.87\linewidth
		]{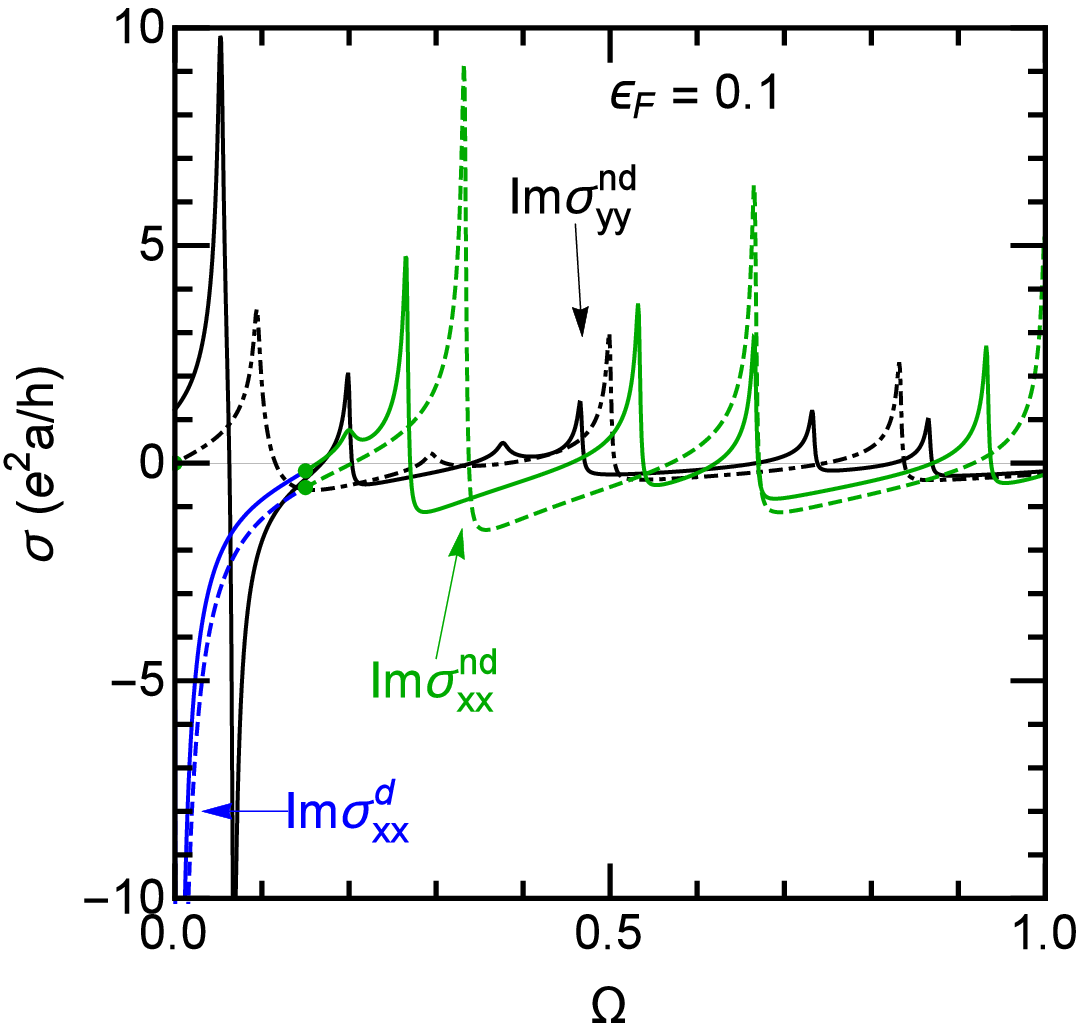}
		\vspace{-0.1cm}
		\caption{As in Fig. \ref{nd1} but for the imaginary part of the total conductivity vs frequency.} %\textcolor{red}{legends?like the previous fig}.} %The curves marking and the other parameters are the same as in Fig. \ref{nd1}.  }
		\label{nd2}
	\end{figure}
	%%%%%%%
	%
	with %{\color{blue} a maximum cutoff %$k_{m}= (
	$\varepsilon_{m}=(k_{m}^2+k_{yn}^{2})^{1/2}$ and $k_m$  the maximum value of $k_x$ below for which the ${\bf k\cdot p}$ theory is valid. %} % is introduced to perform the integration over $k_{y}$ for obtaining the better convergence}, 
	Further, $ v_{x -+ k_{x}}^{n n} v_{x +- k_{x}}^{nn}= v_{x +- k_{x}}^{nn} v_{x -+ k_{x}}^{n n}  =  v_{F}^{2} k_{yn}^{2}/\varepsilon^{2} $ [see Eq. (\ref{v2})], $C_{\pm}=  \Gamma ((2 \hslash v_{F} \varepsilon \pm  \hslash \omega)^{2} + \Gamma^{2})^{-1}$, and  $R_{\pm}= (2 \hslash v_{F} \varepsilon \pm \hslash \omega) ((2 \hslash v_{F} \varepsilon \pm  \hslash \omega)^{2} + \Gamma^{2})^{-1}$. In the limit $k_{yn}=0$, the real and imaginary parts of the nondiagonal conductivity will  vanish as is evident from Eqs. (\ref{v15})-(\ref{v16}). Further, for $\omega=0$ and $\Gamma\neq 0$, the real part  [see Eq. (\ref{v15})] of the nondiagonal conductivity survives whereas the imaginary one vanishes [see Eq. (\ref{v16})]. Also, it can be seen from Eqs. (\ref{v15}) and   (\ref{v16}) that transitions occur between the valence and conduction band with the same %band 
	index $n$.   Some of these transitions are shown schematically in Fig. \ref{tt1} for two values of the Fermi level (dashed red lines) with  $T_{y}^i$ and $T_{x}^i$ denoting the intraband $(i=1)$ and interband $(i=2)$ ones, respectively at the peaks of $\sigma_{xx}^{nd}$ and $\sigma_{yy}^{nd}$.
	
	%{\color{red} 
	For $T=0$ and $E_{F}$ in the gap we have $f_{- k_{x}}^{n}=1$ and $f_{+ k_{x}}^{n}=0$. After evaluating the integrals over $\varepsilon$ in Eqs. (\ref{v15})-(\ref{v16}) 
	%over $\varepsilon$, 
	we rewrite them   in the combined form 
	%%%%%
	\begin{eqnarray}
	\hspace*{-0.6cm}\sigma_{xx}^{nd}(i \omega) & = & \dfrac{i e^{2}U_{+}}{h (\omega_{1}^{2} +  \Gamma_{1}^{2})} \sum_{n} \biggl[ 1+ \dfrac{2 i U_{+} k_{yn}^{2}}{ (\omega_{1}^{2} +  \Gamma_{1}^{2}) p} \ln \dfrac{Q_{+} }{Q_{-}}
	\biggr], \label{v01}
	\end{eqnarray}
	%%%%
	where $\omega_{1} = \hslash \omega/ \hslash v_{F}$, $\Gamma_{1}=\Gamma/ \hslash v_{F}$, %\varepsilon_{0}$,  $\varepsilon_{0}= 2 \pi \hslash v_{F} / a$, 
	$U_{\pm} = (\omega_{1} \pm i \Gamma_{1}) $, $p=(4 k_{yn}^{2}-U_{-}^{2})^{1/2}$, and $Q_{\pm}= p \pm  i U_{-}$. 
	%Further, Eq. (\ref{v01}) is written in combined form for both the real and imaginary parts of conductivity.
	
	For $\mathrm{Re}\sigma_{yy}^{nd} (i\omega)$ we follow the same procedure and from the sum over $n'(\neq n)$, cf. Eq. (\ref{v3}),  we keep only the dominant terms $n'=n\pm1$. %the similar steps as for the calculations of $\sigma_{xx}$, we can find the $\mathrm{Re}\sigma_{yy}^{nd} (i\omega)$ and $\mathrm{Im}\sigma_{yy}^{nd} (i\omega)$
	We then obtain
	%%%
	\begin{eqnarray}
	\notag
	\hspace*{-0.6cm}\mathrm{Re}\sigma_{yy}^{nd}(i\omega) & = &  \dfrac{ e^{2} \hslash  v_{F} }{ h \pi^{2} } \sum_{n} \int_{0}^{k_{m}} dk_{x}
	\notag
	\\ & &
	\times \biggl[ \dfrac{(f_{+}^{n}-f_{+}^{n \pm 1})D_{+}   - (f_{-}^{n}-f_{-}^{n \pm 1})D_{-} }{\varepsilon_{k_{x}}^{n}-\varepsilon_{k_{x}}^{n \pm 1}}  
	\notag
	\\ & &
	+  \dfrac{(f_{+}^{n}-f_{-}^{n \pm 1}) D_{+}   - (f_{-}^{n}-f_{+}^{n \pm 1}) D_{-} }{\varepsilon_{k_{x}}^{n} + \varepsilon_{k_{x}}^{n \pm 1}}     \biggr]
	\label{v17}
	\end{eqnarray}
	%%%%%
	and
	%%%
	\begin{eqnarray}
	\notag
	\hspace*{-0.6cm}\mathrm{Im}\sigma_{yy}^{nd}(i\omega) & = &\dfrac{ e^{2}  \hslash  v_{F} }{ h \pi^{2} } \sum_{n} \int_{0}^{k_{m}} dk_{x}
	\notag
	\\ & &
	\times \biggl[ \dfrac{(f_{+}^{n}-f_{+}^{n \pm 1})E_{+}   +  (f_{-}^{n}-f_{-}^{n \pm 1})E_{-} }{\varepsilon_{k_{x}}^{n}-\varepsilon_{k_{x}}^{n \pm 1}}  
	\notag
	\\ & &
	+  \dfrac{(f_{+}^{n}-f_{-}^{n \pm 1}) E_{+}   +  (f_{-}^{n}-f_{+}^{n \pm 1}) E_{-} }{\varepsilon_{k_{x}}^{n} + \varepsilon_{k_{x}}^{n \pm 1}}     \biggr],\label{v18}
	\end{eqnarray}
	%%%%%
	where $D_{\pm}=  \Gamma (( \hslash v_{F} \varepsilon_{k_{x}}^{n} -  \hslash v_{F} \varepsilon_{k_{x}}^{n \pm 1} \pm  \hslash \omega)^{2} + \Gamma^{2})^{-1}$, and  $E_{\pm}= (\hslash v_{F} \varepsilon_{k_{x}}^{n} + \hslash v_{F} \varepsilon_{k_{x}}^{n \pm 1} \pm \hslash \omega) (( \hslash v_{F} \varepsilon_{k_{x}}^{n}+\hslash v_{F} \varepsilon_{k_{x}}^{n \pm 1} \pm  \hslash \omega)^{2} + \Gamma^{2})^{-1}$ with $ \varepsilon_{k_{x}}^{n}= (k_{yn}^{2}+k_{x}^{2})^{1/2}$.  According to Eqs. (\ref{v17}) and  (\ref{v18}), the absorption occurs between the valence band with index $n$ and the conduction band with $n \pm 1$.
	The integrals over $k_x$ in Eqs. (\ref{v17}) and  (\ref{v18}) are not tractable and we evaluate them numerically. 
	
	In Fig. \ref{nd1} we show   $\mathrm{Re}\sigma_{xx}$ and $\mathrm{Re}\sigma_{yy}$ as   functions of $\Omega$ for $\varepsilon_{F}=0$ (upper panel) and $\varepsilon_{F}=0.1$ (lower panel). The solid curves are for semiconducting nanoribbons $(dm=4)$ and the  dotted ones  for metallic ribbons $(dm=5)$. The optical selection rules $(n-n^{\prime}=\Delta n )$ allow subband index $n$ to change by only $0$ along the $x$ (wire) direction. However, we have $\Delta n =\pm 1$  along the $y$ (confinement) direction, but the amplitude of the peaks is small. Hereafter, we call the transitions satisfying $\Delta n =0$ direct transitions and those  satisfying $\Delta n =\pm 1$  indirect transitions.  In addition, one needs to go from occupied to unoccupied  states through the absorption of photons. The series of peaks corresponding to $\mathrm{Re}\sigma_{xx}$ and $\mathrm{Re}\sigma_{yy}$ occur at $\hslash \omega=-E_{-k_{x}}^{n}+ E_{+k_{x}}^{n}$ and $\hslash \omega=-E_{-k_{x}}^{n}+ E_{+k_{x}}^{n+1}$, respectively. These peaks correspond to the allowed interband transitions in the energy spectrum. The position of the absorption peaks follows the same order as indicated in Fig. \ref{tt1}.  These results for AGNRs are similar to those in Ref. \cite{r4} apart from the contribution $\sigma_{\mu\mu}^{d}(i\omega)$
 which  is completely absent and only the real parts of the conductivities $\sigma_{\mu\mu}^{nd}(i\omega)$ are plotted.
	
	In the upper panel of Fig. \ref{nd1} in which Fermi level is in gap i.e., $\varepsilon_{F}=0$, we can see that a Drude-type intraband transition is allowed in $\mathrm{Re}\sigma_{xx}$ for $dm=5$ due to the nonvanshing $v_{x}$ velocity matrix elements with $\Delta n = 0$ [see   Eq. (\ref{v2}) and transition $D$ in Fig. \ref{tt1}]. On the other hand, we cannot see any type of intraband transitions in  $\mathrm{Re}\sigma_{yy}$ because $v_{y}$ the velocity matrix elements vansh as can be seen from Eq. (\ref{v2}). However, for $dm=4$, only interband absorption transitions are allowed due to the Pauli exclusion principle in both $\mathrm{Re}\sigma_{xx}$ and $\mathrm{Re}\sigma_{yy}$.  But, when we move the Fermi level to $0.1$ [see the red dashed curve in Fig. \ref{tt1}], the  absorption peak, say $T_{x}^{1}$, in $\mathrm{Re}\sigma_{xx}$ is suppressed due to the Pauli exclusion principle  for $dm=4$ in the range $0.017 \leqslant \Omega\leqslant 0.27$ whereas a absorption peak due to intraband  transition $(T_{y}^{1})$ appears in $\mathrm{Re}\sigma_{yy}$ as can be seen in the lower panel of Fig. \ref{nd1}. Moreover, a Drude absorption peaks appear at low $\Omega$ in $\mathrm{Re}\sigma_{xx}$ for both $dm=4$ and $dm=5$. One note worthy feature is that resonance energies $E_{+k_{x}}^{n}-E_{-k_{x}}^{n+1}$ of indirect transitions are appeared between the  $E_{+k_{x}}^{n}-E_{-k_{x}}^{n}$ that are the energies corresponding to absorption peaks of direct transitions.
	
	We have plotted  $\mathrm{Im}\sigma_{xx}$ and $\mathrm{Im}\sigma_{yy}$  versus the dimensionless photon energy $(\Omega)$ in Fig. \ref{nd2}. The absorption peaks in $\mathrm{Im}\sigma_{xx}$ have negative and positive values due to the negative sign between $R_{+}$ and $R_{-}$ terms in Eq. (\ref{v16}), and the peaks corresponding to the transitions $-n \rightarrow n$ and $n \rightarrow -n$ have slightly different energies. This mismatch creates positive and negative peaks in the conductivity. However, the amplitude of the negative peaks is small as compared to that of the  positive ones. This argument applies also to  $\mathrm{Im}\sigma_{yy}$ [see Eq. (\ref{v18})]. 
	%Further, we obtained  intraband transitions peaks similar to those in the real part of the conductivity by changing $\varepsilon_{F} $ from $0$ to $0.1$.
	
	\section{Summary and conclusion}
	
	We studied  dc and ac transport in  both metallic and semiconducting AGNRs. We derived analytical expressions for the diagonal and nondiagonal conductivities by employing  linear response theory. We  found that semiconducting to metallic transitions  occur  by changing the number of rows $(dm)$ [see   Fig. \ref{d1}] in contrast to ordinary waveguides in which   such transitions do not %only the semiconducting ribbons 
	occur, see   Fig. \ref{g1}. In addition, the diagonal conductivity for scattering by screened Coulomb impurities was shown to depend approximately linearly on the carrier density and exhibits  upward cusps when the Fermi level crosses the subbands.
	%peaks to a further linear behavior in the gap among the 
	%These peaks appear when the Fermi level approaches to the subbands. 
	Further, we  showed that the diagonal conductivity varies approximately linearly with the electron concentration in AGNRs, cf.   Fig. \ref{dc1}. % while it is vanished for ordinary waveguide. %(see the Fig. \ref{ord}).
	
	Importantly, in all cases we showed that the  scattering-dependent conductivity is described   quantitatively by a Drude-type contribution $\sigma_{xx}^d(i\omega)$ which, to our knowledge, was not previously reported or explicitly evaluated. We did  show  that this contribution  dominates the response at very low frequencies at which the usual, scattering-independent  contribution near vanishes.
	
	Moreover, we obtained the optical selection rules $\Delta n=0$ along the wire and $\Delta n=\pm 1$  along the confinement direction of AGNRs. We have demonstrated that the peak amplitude of the  indirect transitions   is suppressed contrary to that of the  direct ones. Also, we showed that the absoption of low-energy photons is sensitive to the variation of the Fermi level, %i.e., %intraband transition peaks are appeared when Fermi level is moved from $0$ to $0.1$ [see the Fig. \ref{nd1}] 
	in contrast to  monolayer WSe$_{2}$ \cite{r13}, in which the spectral weight of the interband peaks is continuously redistributed into the intraband ones [see Fig. \ref{nd1}] similar to that of other 2D materials \cite{r14} like graphene, silicene, $\alpha-T_{3}$, and topological insulators. A similar behaviour was found for the imaginary part of the conductivity.   Furthermore, only intraband transitions %along the confinement direction of 
	occur in ordinary waveguides, cf. Fig. \ref{ord} (b) and Eq. (\ref{o3}),  in contrast to AGNRs in which both intra- and inter-band transitions occur [see Figs. (\ref{nd1})-(\ref{nd2}) and Eqs. (\ref{v17})- (\ref{v18})]. %These  findings may be pertinent to the development of  optical devices based on AGNRs.
	
	 The details of the previous %two 
	paragraphs could best be tested, we think, by optical experiments in AGNRs and by contrasting their
	 % , for intraband and interband transitions, 
	results with those in  unconfined graphene or other 2D materials and standard waveguides. 
	%For instance, there are no interband transitions in standard waveguides. 
	The peak positions, that are sensitive to the $dm$-dependent energy gap between the subbands, cf. Eq. (3), could be tuned by a careful choice of $dm$  in experiments performed  in the far infrared (IR) range. This could lead to the development of new optical devices, in particular novel IR photodetectors
	based on photon absorption rather than on thermionic emission or tunnelling in arrays of GRNs proposed in Ref. \cite{r99}.
	Moreover, the  %so far neglected %, to our knowledge, 
	scattering-dependent contribution $\sigma_{xx}^d(i\omega)$ to the  power spectrum  should be evident %in %power absorption experiments 
	at very low frequencies at which the other conductivity contributions $\sigma_{\mu\nu}^{nd}(i\omega)$, as well as $\sigma_{yy}^d(i\omega)$ in our case, vanish ($\Gamma=0$) or nearly so ($\Gamma\neq 0$) and  $\sigma_{xx}^d(i\omega)$ dominates the spectrum,  cf. Ref. \cite{r11}. We are not aware of any such experiments but  hope that they will be carried out and also test the selection rules $\Delta n=0$  and $\Delta n=\pm 1$ mentioned above.
	%%%%%%%%%%%%
	\acknowledgments
	%\vspace{0.2cm}
	M. Z. and P. V. acknowledge the support of the Concordia University Grant No. VB0038 and Concordia University Graduate Fellowship. % The work of M. B. was  supported by ... 
	\appendix
	%\begin{widetext}
	%%%%%%%%%%%%
	\section{relaxation time}
	Within the first Born approximation, the standard formula for relaxation  time takes the form
	%%%%
	\begin{eqnarray}
	\notag
	\dfrac{1}{\tau_{\zeta}}=\dfrac{1}{\tau_{\eta k_{x}}^{n}}&=&\dfrac{2 \pi n_{i}}{\hslash L_{x}} \sum_{n^{\prime}, \eta^{\prime}, k_{x}^{\prime}}  \vert \left\langle n, \eta, k_{x}  \right\vert  U_{x} \left\vert  n^\prime, \eta^{\prime}, k_{x}^{\prime} \right\rangle \vert^{2}\\* 
	&\times&\delta(E_{\eta k_{x}}^{n}-E_{\eta^{\prime}k_{x}^{\prime}}^{n^{\prime}}) (1-\cos \theta), \label{t1}
	\end{eqnarray}
	%%%%
	where $U_{x}=U(x)$ is the impurity potential, $ n_{i}$  the impurity density, and $\theta$  the angle between the initial $(k_{yn},k_{x})$ and final $(k_{yn}^{\prime},k_{x}^{\prime})$ wave vectors. %Furthermore, it is understood that $k_{x}$ is a Fermi wave vector as the transport properties are determined by the states near to the Fermi energy.   
	Equation~ (\ref{t1}) holds only for the elastic scattering.
	%and under assumption that the relaxation time depends only on the energy of state \cite{r5}. 
	The results for two types of impurity potentials are as follows.
	
	{\it  Long-range impurities:} For screened, Coulomb-type impurities  we consider the model potential \cite{r15}
	\begin{equation}
	U_{x}=U_{0} %\dfrac{
	e^{-k_{s} \vert x \vert}\big/\sqrt{\vert x \vert}, \label{t2}
	\end{equation}
	where $U_{0}=2\pi e^{2} \sqrt{c}/\epsilon_{0}\epsilon_{r}$, $k_{s}$  is the screening wave vector, $\epsilon_{0}$   the free space permittivity, $\epsilon_{r}$  the static dielectric constant, and $c$ is the constant of order $1$ in units of inverse length. In this case, 
	we write $U_{x}=\sum_{q_x} U_{q_x} e^{iq_xx}$  with $U_{q_x}= U_0\big\{\big[k_s+\sqrt{k_{s}^{2}+q_x^{2}}\,\big]/(k_{s}^{2}+q_x^{2})\big\}^{1/2}$ the Fourier transform of $U_{x}$. We obtain 
	\begin{eqnarray}
	\notag
	\hspace*{-.6cm} \vert \left\langle n,\eta, k_{x} \right\vert  e^{iq_xx} \left\vert  n^\prime,\eta^\prime,k_{x}^\prime \right\rangle \vert^{2}& = &%\\*=&&
	% \sum_{q} 
	\vert \eta \eta^{\prime} + e^{-i\varphi}\vert^2  \vert U_{q_x} \vert^{2}
	\notag
	\\ & &
	\times \delta_{n, n^{\prime}} \delta_{k_x^\prime+q_x, k_x},
	\end{eqnarray}
	%and 
	%%%%%%%%%
	%\begin{equation}
	%\hspace*{-0.5cm}\vert \left\langle n,\eta, k_{x} \right\vert  U_{x} \left\vert  n^\prime,\eta^\prime,k_{x}^\prime \right\rangle \vert^{2}= \sum_q4 \pi \dfrac{k_{s}+\sqrt{k_{s}^{2}+q^{2}}}{k_{s}^{2}+q^{2}}, \label{t3}
	%\end{equation}
	%%%%%%%%%%%%
	with $ q_{x}= k_{x}-k_{x}^{\prime}$ and $\varphi=\theta_{k_{yn}, k_x}-\theta_{k_{yn}^\prime, k_x^\prime}$. The integration over $q_x$ is straightforward. That over $k_{x}^{\prime}$ is carried out using the properties of the $\delta$ function and only the root $k_{x}^{\prime}=-k_{x}$ of the equation $E_{\eta k_{x}}^{n}-E_{\eta^{\prime}k_{x}^{\prime}}^{n^{\prime}} = 0$ contributes to the integral. For simplicity we also take $\varphi\approx 0$ and %notice that only $\eta=\eta'$ contributes and we 
	use $\theta\approx\pi$. With $k_x$ evaluated at the Fermi level the final result is
	%%%%%%%%%%%
	\begin{eqnarray}
	\hspace*{-0.3cm}\dfrac{1}{\tau_{F}} & = &  \dfrac{  n_{i}U_{0}^{2}}{2  \hslash^2 v_{F} } \dfrac{\sqrt{k_{yn}^{2}+\vert k_{F} \vert^{2}} \,(k_{s}+
		k_{sF})% \sqrt{k_{s}^{2}+4\vert k_{F} \vert^{2}})
	}{\vert k_{F}\vert\, k_{sF}^2}%\vert(k_{s}^{2}+4\vert k_{F} \vert^{2})}
	\label{t4} , 
	\end{eqnarray}
	%%%%
	where $k_{sF}^2=k_{s}^{2}+4\vert k_{F} \vert^{2}$. The term $k_{yn}$ in Eq. (\ref{t4}) denotes the Fermi wave vector for the $n$th subband. %The wave vector of the last subband is small and contributes the most to the value of $1/\tau_{F}$, so that it can, in practice, be approximated by a single term $n=N_{occ}$. 
	In the limit $k_{yn}= 0$  Eq. (\ref{t4}) becomes
	%%%%
	\begin{align}
	& \dfrac{1}{\tau_{F}} = \dfrac{  n_{i}U_{0}^{2}}{2  \hslash^2 v_{F} }  \dfrac{k_{s}+k_{sF}% \sqrt{k_{s}^{2}+4\vert k_{F} \vert^{2}}
	}{k_{sF}^2}%k_{s}^{2}+4\vert k_{F} \vert^{2}
	\label{t5}. 
	\end{align}
	%%%%
	%where $\sqrt{k_{s}^{2}+4\vert k_{F} \vert^{2}}$. 
	Further, for  $k_{s}\gg k_{F}$,  Eq. (\ref{t5})  reduces to
	%%%%
	\begin{align}
	& \dfrac{1}{\tau_{F}} = \dfrac{  n_{i}U_{s}^{2}}{  \hslash^2 v_{F} k_{s} }.
	\label{t6}  
	\end{align}
	%%%%
	
	{\it  Short-range impurities:} we have $U(x)=U_{0}\delta(x-x_{i})$ with $U_{0}$ the constant strength of potential and $x_{i}$ the position of the impurity. In this case, the matrix element becomes $\vert  \left\langle n,\eta,k_{x} \right\vert  U_{x} \left\vert  n^\prime,\eta^\prime,k_{x}^\prime \right\rangle \vert^{2}=U_{0}^{2}$. This leads to
	%%%%%%%%%%%%
	\begin{eqnarray}
	\dfrac{1}{\tau_{F}} & = &\dfrac{ 2 n_{i}U_{0}^{2}}{\pi  \hslash^2 v_{F} } \dfrac{\sqrt{k_{yn}^{2}+\vert k_{F}\vert ^{2}}}{\vert k_{F}\vert}\,.
	\label{t7}
	\end{eqnarray}
	%%%%
	%%%%
	
	For $k_{yn}=0$ Eq. (\ref{t7})  reduces to 
	%%%%
	\begin{align}
	\dfrac{1}{\tau_{F}}= \dfrac{ 2 n_{i}U_{0}^{2}}{\pi  \hslash^2 v_{F} }\, . 
	\label{t8}
	\end{align}
	%%%%
	%%%%
	
	%\end{widetext}


\begin{thebibliography}{99}
		
		
		\bibitem{r1} L. Brey and H. A. Fertig, Phys. Rev. B {\bf 75}, 125434 (2007); 
		H. Zheng, Z. F. Wang, Tao Luo, Q. W. Shi,  and J. Chen, {\it ibid.} {\bf 75}, 044710 (2011);
		
		\bibitem{r2} K.-I. Sakaki, K. Wakabayashi, and T. Enoki, J. Phys. Soc. Jpn. {\bf 80}, 044710 (2011); M.-F. Lin and F.-L. Shyu, {\it ibid.} {\bf 69}, 3529 (2000);
		K. Gundra and A. Shukla, Phys. Rev. B {\bf 83}, 075413 (2011).
		
		\bibitem{r22} Ken-ichi Sasaki, K. Kato, Y. Tokura, K. Oguri, and T. Sogawa, Phys. Rev. B {\bf 84}, 085458 (2011).
		
		\bibitem{r3} L. Brey and H. A. Fertig, Phys. Rev. B {\bf 73}, 235411 (2006); K. O. Wedel, N. A. Mortensen, K. S. Thygesen,  and M. Wubs, Phys. Rev. Lett. {\bf 98}, 155412 (2018); M. Bahrami and P. Vasilopoulos, Optics Express {\bf 25}, 16840 (2017).
		%A. H. Castro Neto, F. Guinea, N. M. R. Peres, K. S. Novoselov, and A. K. Geim, Rev. Mod. Phys. {\bf 81}, 109 (2009).
		
		\bibitem{r4} Y. Ominato and M. Koshino, Phys. Rev. B {\bf 85}, 165454 (2012);  Solid State Commun. {\bf 175-176}, 51 (2013).
		
		\bibitem{nr1} %K. Wakabayashi, M. Fujita, H. Ajiki and M. Sigrist, Phys. Rev. B {\bf 59}, 8271 (1999);  
		J. Liu, Z. Ma, A. R. Wright, and C. Zhang, J. Appl. Phys. {\bf 103}, 103711 (2008).
		
		\bibitem{r5} Li Yang, M. L. Cohen, and S. G. Louie, Nano Lett. {\bf 7}, 3112 (2007). 
		
		\bibitem{r6} K. Wakabayashi, Y. Takane, M. Yamamoto, and M. Sigrist, New J. Phys. {\bf 11}, 095016 (2009).
		
		\bibitem{r7}  Thomas Aktor, Antti-Pekka Jauho, and Stephen R. Power, Phys. Rev. B {\bf 33}, 035446 (2016); J. Guo,  D. Gunlycke, and C. T. White, Appl. Phys. Lett. {\bf 92}, 163109  (2008); P. Hawkins, M. Begliarbekov, M. Zivkovic, S. Strauf, and C. P. Search, J. Phys. Chem. C {\bf 116}, 18382 (2012).
		
		\bibitem{r8} J. Lan,\ J.-S. Wang,  C. K. Gan,  and S. K. Chin, Phys. Rev. B {\bf 79}, 115401 (2009); Ming-Xing Zhai and Xue-Feng Wang, Sci. Rep. {\bf 6}, 36762 (2016).
		
		\bibitem{r88}Ting-Ting Wu,  Xue-Feng Wang, Ming-Xing Zhai,  Hua Liu,  Liping Zhou, and Yong-Jin Jiang, Appl. Phys. Lett. {\bf 100}, 052112 (2012).
		
		\bibitem{r9}  O. Grning, S. Wang, X. Yao, C. A. Pignedoli, G. B. Barin, C. Daniels,  A. Cupo, V. Meunier, X. Feng, A. Narita, K. M\'ullen, P. Ruffieux, and R. Fasel, Nature {\bf 560}, 209 (2018);
		G. Z. Magda, X. Jin, I. Hagymasi, P. Vancso, Z. Osvath, P. Nemes-Incze, C. Hwang, L. P. Biro, and L. Tapaszto,  Nature {\bf 514}, 608 (2014); M. Y. Han,  J. C. Brant, and P. Kim, Phys. Rev. Lett. {\bf 104}, 056801 (2010).
		
		
		\bibitem{r10} M. Charbonneau, K. M. Van Vliet, and P. Vasilopoulos, J. Math. Phys. {\bf 23}, 318 (1982).
		
		\bibitem{r11} V. Vargiamidis, P. Vasilopoulos and G-Q. Hai, J. Phys.: Condens. Matter {\bf 26}, 345303  (2014);   V. Vargiamidis, and P. Vasilopoulos, J. Appl. Phys. {\bf 116}, 063713 (2014).
		
		\bibitem{nr11} S. Zhang, R. Liang, E. Zhang, L. Zhang, and Y. Liu, Phys. Rev. B {\bf 73}, 155316 (2006).
		
		\bibitem{r12} T. Stauber, N. M. R. Peres, and F. Guinea, Phys. Rev. B {\bf 76}, 205423 (2007);
		K. Nomura and A. H. MacDonald, Phys. Rev. Lett. {\bf 98}, 076602 (2007). 
		
		\bibitem{r13} M. Tahir and P. Vasilopoulos, Phys. Rev. B {\bf 94}, 045415 (2016).
		
		\bibitem{r14}  Z. Li and J. P. Carbotte, Phys. Rev. B {\bf 88}, 045414 (2013); %C. J. Tabert and E. J. Nicol, {\it ibid.} 88, 085434 (2013); 
		E. Illes and E. J. Nicol, {\it ibid.} {\bf 94}, 125435 (2016).
		
		\bibitem{r15} P. Vasilopoulos and  F.M. Peeters, Phys. Rev. B {\bf 40}, 10079 (1989).  
		
		\bibitem{r99} V. Ryzhii, M Ryzhii, N. Ryabova, V. Mitin, and T. Otsuji, Jpn. J. Appl. Phys. {\bf 48},  04C144 (2009). 
		
	\end{thebibliography}
\end{document}